\documentclass[prl,twocolumn,showpacs,superscriptaddress,preprintnumbers,nofootinbib,
amsmath,amssymb]{revtex4-1}

\usepackage[colorlinks=true,urlcolor=blue,linkcolor=blue,citecolor=blue]{hyperref}
\usepackage[normalem]{ulem}
\usepackage[capitalize]{cleveref}
 
\usepackage{comment}
\usepackage{flushend}

\usepackage[T1]{fontenc} 

\usepackage{mathtools}

\usepackage{xcolor}
\usepackage{graphicx}
\usepackage{dcolumn}
\usepackage{bm}
\usepackage{multirow}

\usepackage{epstopdf}

\newcommand{\Beq}{\begin{equation}\begin{aligned}}
\newcommand{\Eeq}{\end{aligned}\end{equation}}

\begin{document}

\preprint{IPMU22-0016, KEK-QUP-2023-0008, KEK-TH-2518, KEK-Cosmo-0311}

\title{Enhanced Gravitational Waves from Inflaton Oscillons}

\author{Kaloian D. Lozanov}
\email{klozanov@illinois.edu}
\affiliation{Illinois Center for Advanced Studies of the Universe \& Department of Physics, University of Illinois at Urbana-Champaign, Urbana, IL 61801, USA.}
\affiliation{Kavli Institute for the Physics and Mathematics of the Universe (WPI), UTIAS
The University of Tokyo, Kashiwa, Chiba 277-8583, Japan.}
\author{Volodymyr Takhistov}
\email{vtakhist@post.kek.jp}
\affiliation{International Center for Quantum-field Measurement Systems for Studies of the Universe and Particles (QUP, WPI),
High Energy Accelerator Research Organization (KEK), Oho 1-1, Tsukuba, Ibaraki 305-0801, Japan}
\affiliation{Theory Center, Institute of Particle and Nuclear Studies (IPNS), High Energy Accelerator Research Organization (KEK), Tsukuba 305-0801, Japan
}
\affiliation{Kavli Institute for the Physics and Mathematics of the Universe (WPI), UTIAS
The University of Tokyo, Kashiwa, Chiba 277-8583, Japan.}
\date{\today}

\begin{abstract}
In broad classes of inflationary models the period of accelerated expansion is followed by fragmentation of the inflaton scalar field into localized, long-lived and massive oscillon excitations.
We demonstrate that matter-dominance of oscillons, followed by their rapid decay, significantly enhances the primordial gravitational wave (GW) spectrum. These oscillon-induced GWs, sourced by second-order perturbations, are distinct and could be orders of magnitude lower in frequency than the previously considered GWs associated with oscillon formation. We show that detectable oscillon-induced GW signatures establish direct tests independent from cosmic microwave background radiation (CMB) for regions of parameter space of monodromy, logarithmic and pure natural (plateau) potential classes of inflationary models, among others. We demonstrate that oscillon-induced GWs in a model based on pure natural inflation could be directly observable with the Einstein Telescope, Cosmic Explorer and DECIGO. These signatures offer a new route for probing the underlying inflationary physics. 
\end{abstract}

\maketitle

{\it Introduction}.--
Understanding the earliest stages of the history of the Universe is a fundamental quest of physics. Since the dynamics of the early Universe are associated with high energy scales far beyond the reach of terrestrial laboratories, they are inherently challenging to probe. 

An early period of rapid cosmic inflationary expansion resolves variety of conceptual problems and provides an attractive framework for explaining the observed Universe from initial conditions~\cite{Guth:1980zm,Albrecht:1982wi,Linde:1981mu,Mukhanov:1981xt,Sasaki:1986hm,Sasaki:1995aw}, such as formation of galaxies. Inflation predicts a characteristic nearly scale-invariant power spectrum, described by a spectral index $n_s$,  of the angular anisotropies in the cosmic microwave background (CMB) temperature as well as potentially a non-negligible $B$-mode polarization and a related ratio of tensor-to-scalar fluctuations $r$~\cite{Mukhanov:1990me}. Recent CMB observations established accurate measurements of the scale dependence of the angular power spectrum and a tight upper bound on $r$, strongly disfavoring single-parameter single-field models of the slow-roll inflation~\cite{Planck:2018jri}. Nevertheless, many classes of two-parameter models are fully consistent with the observational constraints. However, lacking variety of additional signatures these models are difficult to test with the CMB alone in the near future. 

Similarly, the possibilities for probing the subsequent stage of post-inflationary reheating \cite{Kofman:1997yn,Allahverdi:2010xz,Lozanov:2019jxc} are restricted. For most reheating scenarios, the only signature is a high-frequency stochastic gravitational wave (GW) background, sourced by the non-linear evolution of matter on sub-horizon scales \cite{Bassett:2005xm,Amin:2014eta}. Such signatures, however, reside outside the detection range of the current and planned GW observatories~\cite{Shandera:2019ufi,LISACosmologyWorkingGroup:2022kbp}.

Stochastic GWs induced by curvature perturbations at second order constitute an attractive window into the workings of the early Universe~(e.g.~\cite{Ananda:2006af,Baumann:2007zm,Saito:2008jc,Inomata:2019ivs,Inomata:2019zqy,Inomata:2020xad,Domenech:2020kqm,Sugiyama:2020roc,Adshead:2021hnm,Domenech:2021wkk,Domenech:2021ztg}).
Induced GWs entail the non-linear sourcing of tensor metric perturbations (i.e. GWs) by scalar metric perturbations on sub-horizon scales after the end of inflation. Typically, induced GWs have been considered in inflationary scenarios involving the generation of an amplified non-scale-invariant feature in the scalar power on small-scales during the slow-roll phase, whose subsequent horizon re-entry and growth after inflation leads to the production of induced GWs (e.g.~\cite{Domenech:2021ztg}). 

In this work we put forth oscillon-induced gravitational waves (OIGWs) as a novel observational signature of the experimentally-favoured two-parameter models of inflation and their post-inflationary dynamics.
In multi-parameter inflationary models, such as pure nature (plateau) inflation \cite{Nomura:2017ehb,Nomura:2017zqj}, monodromy inflation \cite{Silverstein:2008sg,McAllister:2014mpa}, natural inflation~\cite{Adams:1992bn,Freese:1990rb} and hilltop inflation \cite{Antusch:2016con,Antusch:2017flz}, the Universe generically undergoes a long post-inflationary matter-dominated state of expansion.
The matter-dominated state is often associated with the presence of oscillons, long-lived solitonic-like configurations of the real scalar inflaton field, formed through inflaton fragmentation after the period of inflation~\cite{Bogolyubsky:1976nx,Gleiser:1993pt,Kasuya:2002zs,Copeland:1995fq,Broadhead:2005hn,Amin:2010xe,Amin:2010dc,Amin:2011hj}. 
During this period the inflaton condensate oscillates about the minimum of its potential and resonant instabilities in its spatial inhomogeneities can lead to backreaction and oscillon production. 
The violent oscillon-formation process also sources high-frequency, GHz-scale, stochastic GW background \cite{Lozanov:2019ylm,Zhou:2013tsa,Antusch:2016con,Liu:2017hua,Amin:2018xfe,Hiramatsu:2020obh}. Inflaton oscillons could also lead to formation of primordial black holes~\cite{Cotner:2018vug,Cotner:2019ykd}. 
\begin{figure*}[t]
    \centering
\includegraphics[width=0.85\columnwidth]{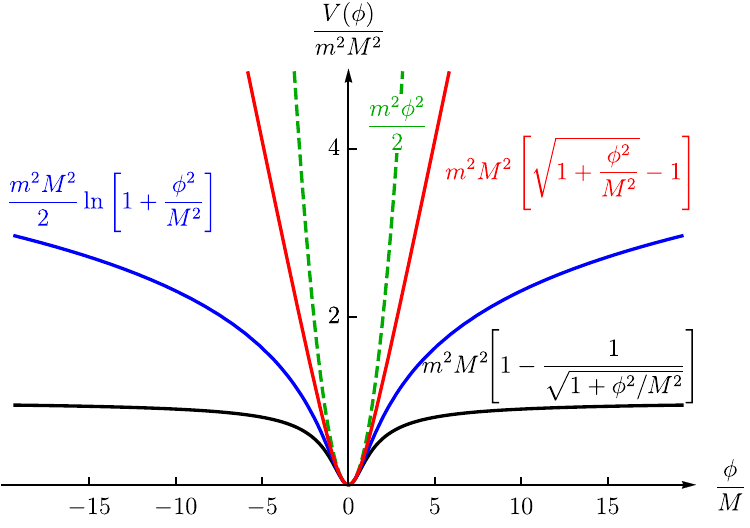} \hspace{4em}
\includegraphics[width=0.85\columnwidth]{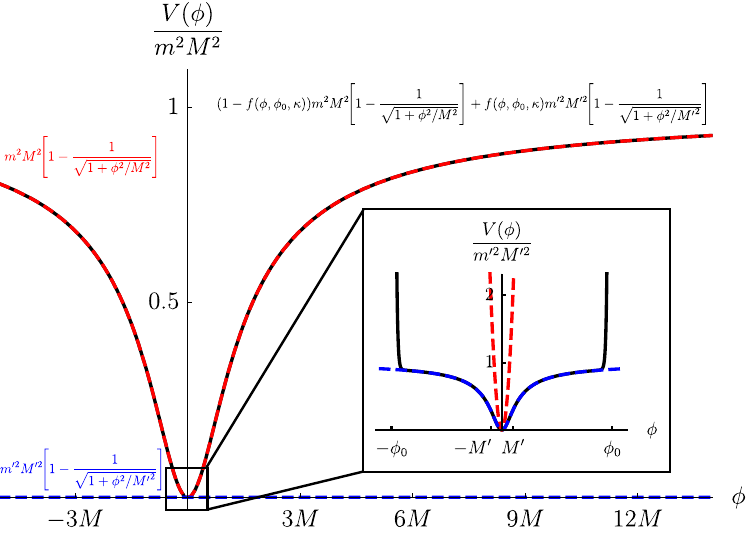}
    \caption{ [Left] Potentials for Eq.~\eqref{eq:monodromy}, log-normal Eq.~\eqref{eq:lognormal} and pure natural (plateau) Eq.~\eqref{eq:plateau} models. [Right]
    Potential of Eq.~\eqref{eq:frank} based on pure natural (plateau)-inflation that exhibits long-lived oscillons, described by model parameters $\kappa\gg1$, $M'\ll\phi_0\ll M$, $m'\ll m$.}
    \label{fig:potentials}
\end{figure*}

We demonstrate that with subsequent rapid oscillon decay the epochs of single-field two-parameter inflation models associated with oscillons will result in dramatically enhanced scale-invariant scalar perturbations and source novel OIGWs signatures. These OIGW signatures are distinct from and can be orders of magnitude lower in frequency than GWs associated with oscillon-formation.
The OIGW signal could lie in the observational range of upcoming GW experiments. Further, we demonstrate that the OIGWs provide a complementary probe of inflationary models that are challenging to test by CMB measurements.

{\it Inflation}.---  Consider single-field models of inflation in which the inflaton, $\phi$, is minimally coupled to gravity, i.e.,
\Beq \label{eq:action}
S=\int d^4x\sqrt{-g}\left[-\frac{M_{\rm Pl}^2}{2}R+\frac{(\partial\phi)^2}{2}-V(\phi)\right]\,
\Eeq
where $R$ is the Ricci scalar, $g$ is the determinant of the metric and $M_{\rm Pl} = 2.435 \times 10^{18}$~GeV is the reduced Planck mass. The scalar inflaton field $\phi$ drives the inflationary expansion of the Universe during its slow-roll phase.

Measurements of the angular anisotropies in the CMB place significant constraints on the shape of the inflaton potential, $V(\phi)$. Observables related to CMB are directly linked to the inflaton potential~\cite{Planck:2018jri}, including the scalar power spectrum amplitude, $A_{\rm s}$, scalar spectral index, $n_{\rm s}$, tensor-to-scalar power ratio, $r$, and the number of expansion $e$-folds before the end of the slow-roll inflation, $N_\star$, when the pivot scale $k_\star$=0.05~Mpc$^{-1}$ exited the horizon (see Supplemental Material~\cite{supmat} for overview).

CMB observations \cite{Planck:2018jri} favor inflaton potentials, $V(\phi)$, having a plateau region away from the minimum (along which the slow-roll inflation occurs), and a power-law minimum (about which the inflaton condensate oscillates at the beginning of reheating). Motivated target potentials consistent with observations include:
\begin{itemize}
  \item {\it Monodromy - } string and supergravity scenarios often lead to power-law models, central realization being axion monodromy \cite{Silverstein:2008sg,McAllister:2014mpa}  
 \begin{equation} \label{eq:monodromy}
     V(\phi) = m^2M^2 \left[\sqrt{1+\dfrac{\phi^2}{M^2}} - 1\right]
 \end{equation}
  \item {\it Pure Natural (Plateau) - } broad classes of axion models, incited by considerations from fundamental theory, lead to universal predictions in $(n_s, r)$ plane described by a converging attractor \cite{Nomura:2017zqj,Nomura:2017ehb}
 \begin{equation} \label{eq:plateau}
     V(\phi) = m^2M^2 \left[1-\frac{1}{\sqrt{1+\phi^2/M^2}}\right]
 \end{equation}
 \item {\it Log-normal - } logarithmic terms in the potential can naturally appear from radiative corrections and supersymmetric theories (e.g.~\cite{Dvali:1994ms})
 \begin{equation} \label{eq:lognormal}
     V(\phi) = \frac{m^2M^2}{2} \ln\left[1+\frac{\phi^2}{M^2}\right]
 \end{equation}
 \end{itemize}

We display the resulting potentials in Fig.~\ref{fig:potentials}.
In Supplemental Material~\cite{supmat} we illustrate the behavior of these potentials in field space relative to quadratic and derive expressions for their observables $A_{\rm s}$, $n_{\rm s}$, $r$ and $N_\star$.

{\it Inflaton oscillons}.---Inflationary potentials shallower than quadratic, such as in classes of models based on monodromy inflation Eq.~\eqref{eq:monodromy}, log-normal inflation Eq.~\eqref{eq:lognormal} and plateau inflation Eq.~\eqref{eq:plateau}, can naturally result in scalar field fragmentation and oscillon formation after inflation (e.g.~\cite{Amin:2010dc,Amin:2010xe}). 
Oscillons are quasi-stable, approximately spherical localized and massive field configurations.
Formation of oscillons during non-linear behavior of the field has been explicitly confirmed by numerous lattice simulations (e.g.~\cite{Amin:2011hj,Lozanov:2019ylm,Hiramatsu:2020obh}).

After the end of inflation, the inflaton oscillates around the minimum of its potential $V$. The non-linearities associated with the behavior of the potential lead to an instability in the fluctuations and  non-adiabatic production of
inflaton particles of definite co-moving momentum,
preheating~\cite{Kofman:1997yn}. The field oscillations by the homogeneous condensate $\overline{\phi}$ can be described by the linearized equations of motions for the perturbations $\phi(t, \bold{x}) = \overline{\phi}(t) + \delta \phi(t, \bold{x})$, which in the static Universe approximation and ignoring metric perturbations are given in Fourier space as
\begin{equation}  \label{eq:lineom}
\partial_t^2 \delta \phi_k + \Big[k^2 + \partial_{\overline{\phi}}^2 V(\overline{\phi}) \Big] \delta \phi_k = 0~.
\end{equation}
Since $\partial_t^2 \overline{\phi} + \partial_{\overline{\phi}} V = 0$, $\partial_{\overline{\phi}}^2 V$ is periodic in time.

The individual momentum modes and their resonant stability can be analyzed with Floquet theory   (e.g.~\cite{Amin:2010xe,Amin:2011hj}), with the general solution to Eq.~\eqref{eq:lineom} being
\begin{equation}
\label{eq:FloqSolution}
    \delta\phi_k(t) = P_+(t) e^{\mu_k t} + P_-(t)e^{-\mu_k t}~,
\end{equation}
where $P_{\pm}(t)$ are periodic functions determined by the initial conditions, $\mu_k$ is the Floquet exponent. The magnitude of the real part of the Floquet exponent, $|{\Re}(\mu_k)|$, characterises the exponential mode growth, provided it is sufficiently faster than the Hubble expansion rate $H \sim t^{-1}$, after inflation. 
In Supplemental Material~\cite{supmat} we explicitly perform instability analysis, compute Floquet charts and find instability bands for the monodromy, log-normal and pure natural (plateu) potentials.

Unstable modes lead to parametric resonance and copious formation of oscillons, for potentials  shallower than quadratic with self-interactions~(e.g.~\cite{Lee:1991ax,Amin:2011hj,Amin:2013ika}). 
The resulting oscillons have high density contrast $\delta = \rho/\overline{\rho} > 1$ and comoving number density $a^3 n_{\rm sol} \sim (k_{\rm nl}/2 \pi)^3$ determined by the dominant non-linear wave mode $k_{\rm nl}$. They oscillate at the approximate frequency of the effective mass $m$, with typical size $R \sim 10/m$ that is much smaller than the relevant horizon scale.

Although oscillons are not protected by a topological invariant and a conserved charge they can be remarkably long lived, which from semi-analytic studies can be attributed to an approximate adiabatic invariance of the oscillating field~\cite{Kasuya:2002zs,Mukaida:2016hwd,Ibe:2019vyo}. Oscillon lifetime has been recently systematically explored by detailed numerical calculations for broad classes of potentials~\cite{Zhang:2020bec,Olle:2019kbo,Antusch:2019qrr}, including large amplitude oscillons, confirming that oscillons are generically long-lived, with a lifetime in some models (including the ones considered here) exceeding $\Delta t \gtrsim 6\times10^8/m$ followed by a rapid decay. Since oscillons behave as presureless dust, they can naturally lead to long periods of (oscillon)matter-domination. 

From instability analysis, we find that oscillons form soon after the end of slow-roll inflation provided $M\lesssim10^{-2}M_{\rm Pl}$ (see Supplemental Material~\cite{supmat} and discussion after Eq. \eqref{eq:FloqSolution}). The mean energy density of the Universe at the time of formation is $\rho_{\rm f}\sim m^2M^2$,
and at the time of decay
$\rho_{\rm d}\sim (m/n)^2M_{\rm Pl}^2$. $n=m\Delta t/(2\pi)$ is the number of oscillations it takes for the oscillons to decay and we assume $H_{\rm d}\sim m/n$. Looking ahead, the ratio $\rho_{\rm f}/\rho_{\rm d}\sim n^2M^2/M_{\rm Pl}^2$ determines the duration of production and thereby the magnitude of the induced GW signal. 

The classes of motivated potentials above allow for theories with significantly prolonged oscillon lifetime. A realization of this can be seen within the modified pure natural (plateau) model given by
\Beq \label{eq:frank}
V=&~\,(1-f(\phi,\phi_0,\kappa)) m^2M^2 \left[1-\frac{1}{\sqrt{1+\phi^2/M^2}}\right]\\
&~+f(\phi,\phi_0,\kappa) m'^2M'^2 \left[1-\frac{1}{\sqrt{1+\phi^2/M'^2}}\right]\,,\\
f=&~\,\frac{1}{1+\exp(\kappa(|\phi|/\phi_0-1))}\,.
\Eeq
For $\kappa\gg1$, $M\gg\phi_0\gg M'$, $mM\gg m'M'$, we have for $|\phi|\lesssim \phi_0$ the oscillon potential $V\approx m'^2M'^2(1-1/\sqrt{1+\phi^2/M'^2})$, with $m'\ll m$ and thus a much longer oscillon lifetime, $\Delta t\gtrsim6\times10^8/m'$, than in the basic plateau model of Eq.~\eqref{eq:plateau}. For this small-field plateau potential, our stability analysis for plateau potential still applies, with $m$ and $M$ replaced by $m'$ and $M'$.

{\it Oscillon-induced gravitational waves}.---
Primordial scalar perturbations induce GWs at second order. While often suppressed~\cite{Inomata:2019zqy}, sudden transition from an early matter-dominated era (eMD) to radiation-dominated (RD) era significantly enhances induced GWs~\cite{Alabidi:2013lya,Inomata:2019ivs}. Such amplification of GWs has been extensively explored in the context of evaporating primordial black holes, whose ``gas'' matter-dominates the Universe before suddenly transitioning to radiation due to Hawking evaporation~(e.g.~\cite{Inomata:2019zqy,Domenech:2020ssp,Papanikolaou:2020qtd,Inomata:2020xad,Domenech:2020kqm,Domenech:2021wkk,Domenech:2021ztg}). 
 
After the inflaton fragments into oscillons, they matter-dominate the Universe before rapidly decaying, leading to an onset of reheating. The length of matter-dominated era is set by the oscillon lifetime. We demonstrate that due to the rapid decay of the oscillons at the end of their lifetime, these processes result in significant induced GW production. We stress, however, that in our scenario there is direct connection of induced GWs with the inflaton's dynamics. Oscillon isocurvature perturbations could lead to yet another source of induced GWs~(e.g.~\cite{Passaglia:2021jla,Domenech:2021and}), study of which we leave for future work~\cite{Lozanov:2022fut}. Hence, GW signal estimates in this study are conservative. We note that prolonged transition between eMD and RD eras could result in additional suppression of induced GWs (e.g.~\cite{Inomata:2020lmk,Domenech:2021ztg}). However, for oscillons, once the core amplitude of oscillations reaches a critical value the objects are expected to explosively disperse on negligible timescales $\delta t \ll H_d^{-1}$ compared to Hubble timescales $H_d^{-1}$ associated with oscillon decay, typically differing by orders of magnitude~\cite{Zhang:2020bec}. The exact realization of this depends on the model, e.g. for small-amplitude oscillons coupled to daughter fields see Ref.~\cite{Hertzberg:2010yz}. Our results cover a broad range of scenarios where inflaton couples weakly to the SM sector, as relevant for reheating.

We consider that inflation results in a generic primordial curvature (scalar) power spectrum described by
\begin{equation} \label{eq:powerspec}
    \mathcal{P}_{\zeta} = A_s \Theta(k_{\rm max} - k) \Big( 
    \dfrac{k}{k_{\star}}\Big)^{n_s-1}~,
\end{equation}
where $k_{\rm max}$ is the cutoff scale.

For the calculation of the induced GWs we follow \cite{Kohri:2018awv,Inomata:2019ivs}.
We consider that that eMD of the Universe rapidly transitioned to RD, as is the case for oscillon-dominated phase, at a conformal time $\eta = \eta_{\rm R}$. 

The energy density of induced GWs is
\begin{equation}
    \Omega_{\rm GW}(\eta, k) = \dfrac{1}{24}\Big(\dfrac{k}{a(\eta)H(\eta)}\Big)^2   \overline{\mathcal{P}_h(\eta,k)}~,  
\end{equation}
where $\eta$ is conformal time and $\overline{\mathcal{P}_h(\eta,k)}$ is the time-averaged power spectrum of GWs related to the curvature perturbations as \cite{Inomata:2016rbd,Kohri:2018awv}
\begin{align} \label{eq:powerspec}
    \overline{\mathcal{P}_h(\eta,k)} =&~ 4 \int_0^{\infty} {\rm d}v \int_{|1-v|}^{1+v} {\rm d}u \Big(\dfrac{4v^2 - (1 + v^2 - u^2)^2}{4 v u}\Big)^2 \notag\\
    & \times \overline{I^2 (u,v,k,\eta,\eta_{\rm R})} \mathcal{P}_{\zeta}(uk) \mathcal{P}_{\zeta}(vk)~,
\end{align}
where the kernel function $I(u,v,k,\eta,\eta_{\rm R})$ describes dynamics of scalar and tensor (i.e. GW) perturbations that depends on (using Green's function methods) a source function described by the gravitational potential $\Phi$ and its conformal derivative $\Phi^{\prime}$ (see Supplemental Material~\cite{supmat}).

As the oscillons suddenly decay at the end of the lifetime, $\Phi^{\prime}$ jumps from $\Phi^{\prime} = 0$ in eMD to $\Phi^{\prime} \neq 0$ in RD phase. The resulting significant amplitude increase of $\Phi^{\prime}$ dramatically enhances induced GWs through $I \sim (\Phi^{\prime}/\mathcal{H})^2$~(e.g.~\cite{Inomata:2019ivs}). Physically, the enhancement results from rapid oscillations of perturbations with unsuppressed amplitudes in $\Phi$, constant before $\eta_{\rm R}$, on timescales shorter than the decay.

The approximate general expressions for the induced GWs from the power spectrum of Eq.~\eqref{eq:powerspec} are reviewed in the Supplemental Material~\cite{supmat}. For the case of spectral index $n_s \simeq 1$ the GW spectrum can be obtained from simplified analytic expressions~\cite{Inomata:2019ivs}
\begin{align} 
\frac{\Omega_{\text{GW}}(\eta_c,k)}{A_{\text{s}}^2} \simeq ~~~~~~~~~~~~~~~~~~~~~~~&  \notag\\
\begin{cases}
0.8 & (x_{\text{R}} \lesssim 150 x_{\text{max,R}}^{-5/3}) \\
3 \times 10^{-7} x_{\text{R}}^3 x_{\text{max,R}}^5  &   (150 x_{\text{max,R}}^{-5/3} \lesssim x_{\text{R}} \ll 1) \\
1 \times 10^{-6} x_{\text{R}} x_{\text{max,R}}^5 & (1 \ll x_{\text{R}} \lesssim x_{\text{max,R}}^{5/6}) \\
3 \times 10^{-7}   x_{\text{R}} ^7 &     (x_{\text{max,R}}^{5/6} \lesssim x_{\text{R}}   \lesssim x_{\text{max,R}})  \\
\text{(sharp drop)} & (x_{\text{max,R}} \lesssim x_{\text{R}} \leq 2 x_{\text{max,R}})
 \end{cases},  \label{eq:specGWcase}
\end{align}
where $x_{R} = k \eta_{R}$ and $x_{\text{max,R}} = k_{\text{max}} \eta_R$.

After cosmological evolution considerations, the resulting GW spectrum at present time is given by~(e.g.~\cite{Lozanov:2019ylm})
\begin{equation}
\label{eq:omegagw0}
  \Omega_\text{GW}(\eta_0,k)h^2 = 0.39 \left( \frac{g_{*,\text{c}}}{106.75} \right)^{-1/3} \Omega_{\text{r},0}h^2 \Omega_\text{GW}(\eta_\text{c},k)~,
\end{equation}
where $\Omega_{\text{r},0}h^2 \simeq 4.2\times 10^{-5}$ is the current radiation energy density and $\eta_c$ is the moment when $\Omega_{\rm GW}$ becomes constant after gravitational potential has sufficiently decayed soon after modes re-enter the horizon, well before matter-radiation equality, $x_{\rm R} = k \eta_{\rm R}$. Note however, that for a sudden eMD-RD transition $\eta_c\approx\eta_{\rm R}$. We display our results for conservative realization of $x_{\rm max,R} = k_{\rm max} \eta_{\rm R} \leq 450$ that approximately limits the regimes to linear analysis~(e.g.~\cite{Inomata:2019zqy}), as well as those slightly extrapolated to non-linear regimes by considering integration over additional momentum modes $k$ up to factor of $\sim 2$ larger via $x_{\rm max,R} = k_{\rm max} \eta_{\rm R} \leq 10^3$ (following discussion in e.g~\cite{Assadullahi:2009nf}). We leave detailed numerical and simulation analyses of these effects for each model, which can further amplify our signatures from linear analyses, for future work.

\begin{figure}[t]
    \centering
    \includegraphics[width=3.8in]{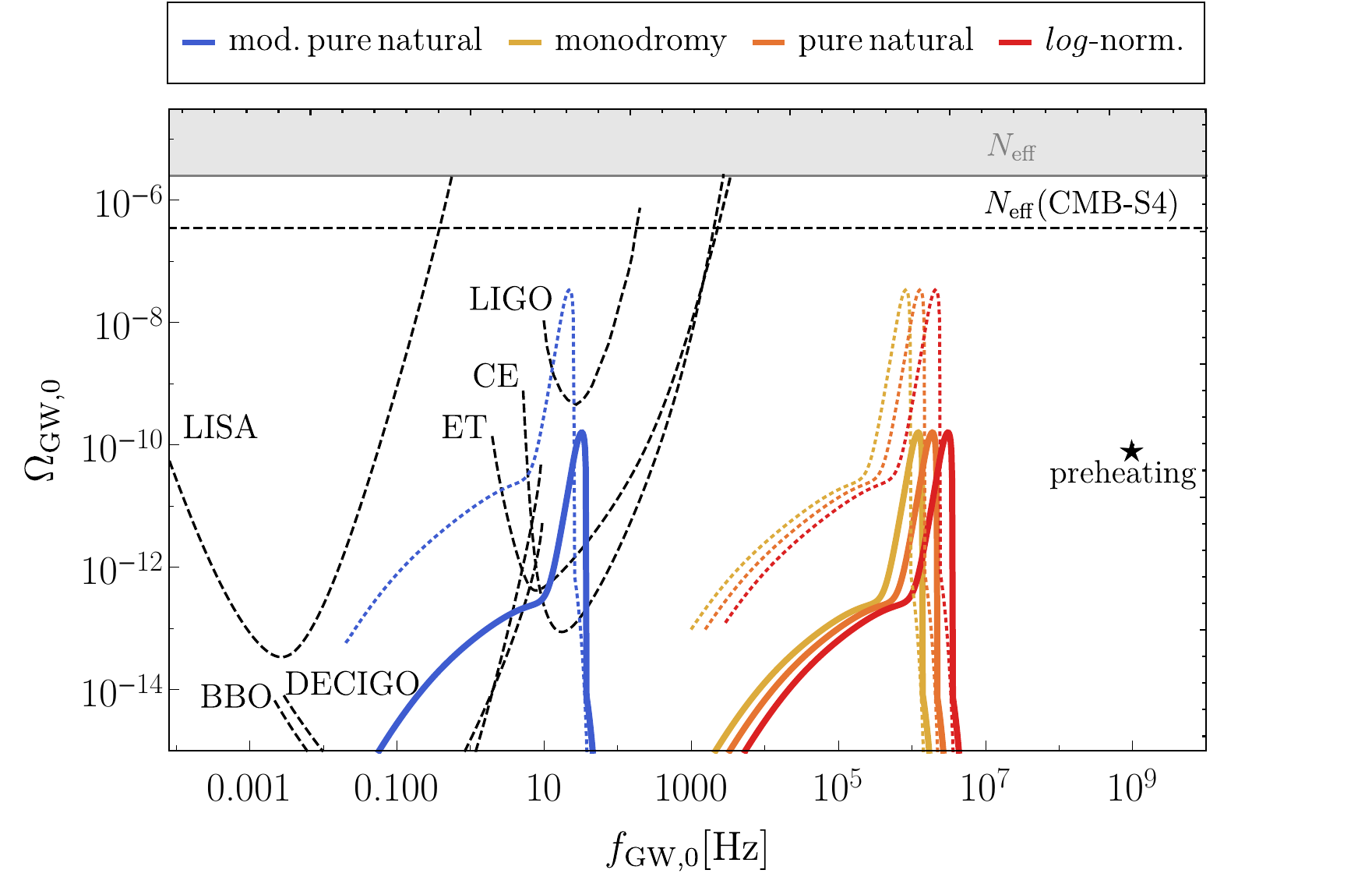}
    \caption{Oscillon-induced gravitational wave signatures. Thick solid curves on the right denote slow-roll inflation with $N_\star =60$ and $M=10^{-2}M_{\rm Pl}$, driven by the monodromy (light orange), $\log$-normal (dark orange) and pure natural plateau-type (red) potentials, with inflaton oscillon life-time described by $n=10^{10}$. The blue solid curve describes modified pure natural (plateau) model with $n=10^{10}, m'=10^{-13} M_{\rm Pl}, M'=10^{-2} M_{\rm Pl}, m=1.5\times10^{-5} M_{\rm Pl}, M=M_{\rm Pl}, \phi_0=10^{-1}M_{\rm Pl}, \kappa=10^2$. The corresponding signatures for longer lived oscillons with $n=10^{11}$, extrapolating slightly into the nonlinear regime, are depicted by the coloured dashed lines. The current $N_{\rm eff}$ constraints from CMB and BBN~\cite{Planck:2018vyg} and expected sensitivity for CMB-S4~\cite{Abazajian:2019eic}, LISA, Einstein Telescope (ET), Cosmic Explorer (CE), Big Bang Observer (BBO) and DECIGO~\cite{Schmitz:2020syl} as well as LIGO O5 \cite{LIGOScientific:2016fpe} are also displayed. We display by a star expected GWs from oscillon formation associated with preheating as discussed in previous literature. }
    \label{fig:limits}
\end{figure}

In our scenarios, at the time of oscillon formation all sub-horizon density perturbations, $k\gtrsim\mathcal{H}_{\rm f}$, collapse into oscillons. During the ensuing oscillon matter-domination era, the modes which undergo gravitational instability are the ones which re-enter the horizon before the oscillon decay, $\mathcal{H}_{\rm f}>k>\mathcal{H}_{\rm d}$. Thus, we set $k_{\rm max}=\mathcal{H}_{\rm f}$. Since matter-radiation equality coincides with the time of oscillon decay, $\eta_{\rm R}=\eta_{\rm d}$, one can show that $x_{\rm max,R}=(\rho_{\rm f}/\rho_{\rm d})^{1/6}\sim (nM/M_{\rm Pl})^{1/3}$. According to Eq. \eqref{eq:specGWcase}, the induced GW signal has a pronounced peak whose magnitude and frequency today are given by $\Omega_{\rm GW,0}\sim {\rm few}\times10^{-29}(nM/M_{\rm Pl})^{7/3}$ and $f_{\rm GW,0}\sim {\rm few}\times10^{10}(m^{3}M^{2}/(nM_{\rm Pl}^{5}))^{1/6}\rm{Hz}$, where to compute the former we used Eq. \eqref{eq:omegagw0} and to estimate the latter we used $f_{\rm GW,0}\sim10^{10}{k_{\rm max}/(a_{\rm d}\rho_{\rm d}^{1/4})}\rm Hz$ \cite{Lozanov:2019ylm}. For the monodromy, plateau and log-normal oscillons, $n$ is only bounded from below ($n\gtrsim10^8$) \cite{Zhang:2020bec}, whereas $M$ is bounded from above\footnote{The classicality of the resonant inflaton dynamics imposes an additional constraint on the model parameters, leading to a lower bound on $M$ \cite{Lozanov:2019ylm}. Oscillon formation, $\rho_{\rm f}\gtrsim \rho_{\rm grad}(k_{\rm nl}/a_{\rm f})$, is classical, when the backreacting gradient energy is of modes of occupancy greater than one, $\rho_{\rm grad}(k_{\rm nl}/a_{\rm f})\gtrsim \rho_{\rm BD}(k_{\rm nl}/a_{\rm f})\sim (k_{\rm nl}/a_{\rm f})^4$, where $\rho_{\rm BD}$ is the energy density in the Bunch-Davies vacuum \cite{Birrell:1982ix}. For $k_{\rm nl}/a_{\rm f}\sim m$, see Supplemental Material~\cite{supmat}, $M\gtrsim m$, implying $M\gtrsim 5.7,13,30\times 10^{-4}M_{\rm Pl}$ for monodromy, plateau and log-normal inflation, respectively. Relaxing the assumptions can modify the allowed range in $M$.} ($M\lesssim10^{-2}M_{\rm Pl}$). In order to remain in the linear approximation, we use the upper bound on $M$ and set $n=10^{10}$, whereas we extrapolate in the nonlinear regime when $n=10^{11}$.

Our prediction for induced GWs holds for reheating scenarios in which the inflaton couplings are sufficiently weaker than its self-interactions. In particular, we consider that the lifetime of the inflaton due to (non-)perturbative decays into daughter particles is longer than the oscillon formation and decay time scales. Hence, the influence on the oscillon dynamics of the other (daughter) fields coupled to the inflaton is negligible. Such non-instant reheating constitutes among the simplest minimal scenarios where oscillons can be realized and very often appears throughout 
oscillon literature, see review~\cite{Amin:2014eta} and e.g.~\cite{Hiramatsu:2020obh,Antusch:2016con}. Moreover, these scenarios are generic, since they are expected for field interactions that do not require significant fine-tuning. For instance, the perturbative inflaton lifetime due to a minimal Yukawa coupling to a light fermion $\psi$, $\mathcal{L}_{\rm int}=\lambda_\psi\phi \bar{\psi}\psi$, is $\Delta t_{\phi\rightarrow\bar{\psi}\psi}\sim 8\pi/(\lambda_\psi^2m)$. Hence, the fermion's impact on oscillon formation and decay is negligible for $\lambda_\psi\ll n^{-1/2}$. This translates to a realistic upper bound of $10^{-4}-10^{-5}$ on Yukawa couplings, which is larger and hence less restrictive than the Standard Model electron coupling. Depending on model details non-perturbative effects could also play a role. However,
in a minimal example with a scalar daughter $\chi$ field coupling of $g m \chi^2 \phi$, the
possible bosonic non-perturbative effects on oscillon dynamics already become 
negligible when $g/\sqrt{\lambda} \lesssim 0.2$, where $\lambda$ is strength of $\phi$ self-interactions~\cite{Hertzberg:2010yz}. More so, for additional fermionic interactions, it is expected that the non-perturbative regime is even further restricted, due to the lack of Bose-enhancement caused by Pauli blocking.

In Fig.~\ref{fig:limits} we display induced GWs computed using the full formalism (see Supplemental Material~\cite{supmat} from inflaton oscillons for the monodromy, log-normal, pure natural (plateau)-inflation as well as modified pure natural (plateau)-inflation model potentials along with existing constraints. We also contrast our novel signatures with expected GWs from oscillon formation associated with preheating as discussed in previous literature (e.g.~\cite{Lozanov:2019ylm,Zhou:2013tsa,Antusch:2016con,Liu:2017hua,Amin:2018xfe,Hiramatsu:2020obh}).
The values of $m$ and $M$ are in agreement with CMB bounds on slow-roll inflation (see Supplemental Material~\cite{supmat} In the monodromy, log-normal and pure natural-inflation models oscillons form when $\bar{\phi}\sim M$, whereas in the modified pure natural inflation when $\bar{\phi}\sim M'$.
We demonstrate that the induced GWs are orders above in amplitude compared to GWs associated with oscillon-formation from preheating-type resonant instabilities \cite{Lozanov:2019ylm}, $f_{\rm GW,0}\sim(M_{\rm Pl}/M)^{1/2}10^8$~Hz, $\Omega_{\rm GW,0}\sim10^{-6}(M/M_{\rm Pl})^2$, (see also \cite{Zhou:2013tsa,Liu:2017hua,Amin:2018xfe,Hiramatsu:2020obh}). 

Pure natural inflation is viable with current CMB constraints at the level of one standard deviation and monodromy as well as log-normal models are at few standard deviations. Thus, an independent test is essential to establish their status. Induced GWs from oscillons could allow to probe these models with CMB-S4~\cite{Abazajian:2019eic} through constraints on $N_{\rm eff}$, in addition to other preheating scenarios~\cite{Adshead:2019lbr,Adshead:2019igv}. 
Further, modified pure natural (plateau) model can be directly testable with Einstein Telescope, Cosmic Explorer and DECIGO at GW frequencies significantly smaller than conventionally associated with oscillon formation (for a review of high frequency GWs see~\cite{Aggarwal:2020olq}). Our results and limits can be readily applied to GW strains, using $h_c(f)=1.26\times10^{-18}(f/{\rm Hz})^{-1}\sqrt{\Omega_{\rm GW,0}h^2}$~\cite{Caprini:2018mtu}.

{\it Conclusions}.---
Generic models of inflation support production of oscillons through inflaton field fragmentation. Current CMB observations favor inflation models with attractive self-interactions, necessary for oscillon formation. We have demonstrated that oscillon matter-domination phase followed by rapid oscillon decay results in dramatic production of induced GWs. These novel signatures are significantly enhanced at lower frequencies compared to GWs associated with oscillon formation previously studied. We have shown that such signatures can provide a crucial independent test for regions of parameter space of broad classes of inflationary models, such as pure natural (plateau), log-normal and monodromy potentials. Induced GWs associated with modified pure natural inflation could be directly observed with the Einstein Telescope, Cosmic Explorer and DECIGO. These results establish a new window into the early Universe dynamics.
 
\acknowledgments

We thank P. Adshead, M. Amin, G. Domenech,
K. Kohri, and M. Sasaki for helpful discussions. K.L.
was supported in part by the NASA Astrophysics Theory
Grant No. NNX17AG48G. V.T. acknowledges support by the
World Premier International Research Center Initiative
(WPI), MEXT, Japan and JSPS KAKENHI Grant No. 23K13109.

\bibliography{BibAuto}

\begin{thebibliography}{81}%
\makeatletter
\providecommand \@ifxundefined [1]{%
 \@ifx{#1\undefined}
}%
\providecommand \@ifnum [1]{%
 \ifnum #1\expandafter \@firstoftwo
 \else \expandafter \@secondoftwo
 \fi
}%
\providecommand \@ifx [1]{%
 \ifx #1\expandafter \@firstoftwo
 \else \expandafter \@secondoftwo
 \fi
}%
\providecommand \natexlab [1]{#1}%
\providecommand \enquote  [1]{``#1''}%
\providecommand \bibnamefont  [1]{#1}%
\providecommand \bibfnamefont [1]{#1}%
\providecommand \citenamefont [1]{#1}%
\providecommand \href@noop [0]{\@secondoftwo}%
\providecommand \href [0]{\begingroup \@sanitize@url \@href}%
\providecommand \@href[1]{\@@startlink{#1}\@@href}%
\providecommand \@@href[1]{\endgroup#1\@@endlink}%
\providecommand \@sanitize@url [0]{\catcode `\\12\catcode `\$12\catcode
  `\&12\catcode `\#12\catcode `\^12\catcode `\_12\catcode `\%12\relax}%
\providecommand \@@startlink[1]{}%
\providecommand \@@endlink[0]{}%
\providecommand \url  [0]{\begingroup\@sanitize@url \@url }%
\providecommand \@url [1]{\endgroup\@href {#1}{\urlprefix }}%
\providecommand \urlprefix  [0]{URL }%
\providecommand \Eprint [0]{\href }%
\providecommand \doibase [0]{http://dx.doi.org/}%
\providecommand \selectlanguage [0]{\@gobble}%
\providecommand \bibinfo  [0]{\@secondoftwo}%
\providecommand \bibfield  [0]{\@secondoftwo}%
\providecommand \translation [1]{[#1]}%
\providecommand \BibitemOpen [0]{}%
\providecommand \bibitemStop [0]{}%
\providecommand \bibitemNoStop [0]{.\EOS\space}%
\providecommand \EOS [0]{\spacefactor3000\relax}%
\providecommand \BibitemShut  [1]{\csname bibitem#1\endcsname}%
\let\auto@bib@innerbib\@empty
\bibitem [{\citenamefont {Guth}(1981)}]{Guth:1980zm}%
  \BibitemOpen
  \bibfield  {author} {\bibinfo {author} {\bibfnamefont {A.~H.}\ \bibnamefont
  {Guth}},\ }\href {\doibase 10.1103/PhysRevD.23.347} {\bibfield  {journal}
  {\bibinfo  {journal} {Phys. Rev. D}\ }\textbf {\bibinfo {volume} {23}},\
  \bibinfo {pages} {347} (\bibinfo {year} {1981})}\BibitemShut {NoStop}%
\bibitem [{\citenamefont {Albrecht}\ and\ \citenamefont
  {Steinhardt}(1982)}]{Albrecht:1982wi}%
  \BibitemOpen
  \bibfield  {author} {\bibinfo {author} {\bibfnamefont {A.}~\bibnamefont
  {Albrecht}}\ and\ \bibinfo {author} {\bibfnamefont {P.~J.}\ \bibnamefont
  {Steinhardt}},\ }\href {\doibase 10.1103/PhysRevLett.48.1220} {\bibfield
  {journal} {\bibinfo  {journal} {Phys. Rev. Lett.}\ }\textbf {\bibinfo
  {volume} {48}},\ \bibinfo {pages} {1220} (\bibinfo {year}
  {1982})}\BibitemShut {NoStop}%
\bibitem [{\citenamefont {Linde}(1982)}]{Linde:1981mu}%
  \BibitemOpen
  \bibfield  {author} {\bibinfo {author} {\bibfnamefont {A.~D.}\ \bibnamefont
  {Linde}},\ }\href {\doibase 10.1016/0370-2693(82)91219-9} {\bibfield
  {journal} {\bibinfo  {journal} {Phys. Lett. B}\ }\textbf {\bibinfo {volume}
  {108}},\ \bibinfo {pages} {389} (\bibinfo {year} {1982})}\BibitemShut
  {NoStop}%
\bibitem [{\citenamefont {Mukhanov}\ and\ \citenamefont
  {Chibisov}(1981)}]{Mukhanov:1981xt}%
  \BibitemOpen
  \bibfield  {author} {\bibinfo {author} {\bibfnamefont {V.~F.}\ \bibnamefont
  {Mukhanov}}\ and\ \bibinfo {author} {\bibfnamefont {G.~V.}\ \bibnamefont
  {Chibisov}},\ }\href@noop {} {\bibfield  {journal} {\bibinfo  {journal} {JETP
  Lett.}\ }\textbf {\bibinfo {volume} {33}},\ \bibinfo {pages} {532} (\bibinfo
  {year} {1981})}\BibitemShut {NoStop}%
\bibitem [{\citenamefont {Sasaki}(1986)}]{Sasaki:1986hm}%
  \BibitemOpen
  \bibfield  {author} {\bibinfo {author} {\bibfnamefont {M.}~\bibnamefont
  {Sasaki}},\ }\href {\doibase 10.1143/PTP.76.1036} {\bibfield  {journal}
  {\bibinfo  {journal} {Prog. Theor. Phys.}\ }\textbf {\bibinfo {volume}
  {76}},\ \bibinfo {pages} {1036} (\bibinfo {year} {1986})}\BibitemShut
  {NoStop}%
\bibitem [{\citenamefont {Sasaki}\ and\ \citenamefont
  {Stewart}(1996)}]{Sasaki:1995aw}%
  \BibitemOpen
  \bibfield  {author} {\bibinfo {author} {\bibfnamefont {M.}~\bibnamefont
  {Sasaki}}\ and\ \bibinfo {author} {\bibfnamefont {E.~D.}\ \bibnamefont
  {Stewart}},\ }\href {\doibase 10.1143/PTP.95.71} {\bibfield  {journal}
  {\bibinfo  {journal} {Prog. Theor. Phys.}\ }\textbf {\bibinfo {volume}
  {95}},\ \bibinfo {pages} {71} (\bibinfo {year} {1996})},\ \Eprint
  {http://arxiv.org/abs/astro-ph/9507001} {arXiv:astro-ph/9507001} \BibitemShut
  {NoStop}%
\bibitem [{\citenamefont {Mukhanov}\ \emph {et~al.}(1992)\citenamefont
  {Mukhanov}, \citenamefont {Feldman},\ and\ \citenamefont
  {Brandenberger}}]{Mukhanov:1990me}%
  \BibitemOpen
  \bibfield  {author} {\bibinfo {author} {\bibfnamefont {V.~F.}\ \bibnamefont
  {Mukhanov}}, \bibinfo {author} {\bibfnamefont {H.~A.}\ \bibnamefont
  {Feldman}}, \ and\ \bibinfo {author} {\bibfnamefont {R.~H.}\ \bibnamefont
  {Brandenberger}},\ }\href {\doibase 10.1016/0370-1573(92)90044-Z} {\bibfield
  {journal} {\bibinfo  {journal} {Phys. Rept.}\ }\textbf {\bibinfo {volume}
  {215}},\ \bibinfo {pages} {203} (\bibinfo {year} {1992})}\BibitemShut
  {NoStop}%
\bibitem [{\citenamefont {Akrami}\ \emph {et~al.}(2020)\citenamefont {Akrami}
  \emph {et~al.}}]{Planck:2018jri}%
  \BibitemOpen
  \bibfield  {author} {\bibinfo {author} {\bibfnamefont {Y.}~\bibnamefont
  {Akrami}} \emph {et~al.} (\bibinfo {collaboration} {Planck}),\ }\href
  {\doibase 10.1051/0004-6361/201833887} {\bibfield  {journal} {\bibinfo
  {journal} {Astron. Astrophys.}\ }\textbf {\bibinfo {volume} {641}},\ \bibinfo
  {pages} {A10} (\bibinfo {year} {2020})},\ \Eprint
  {http://arxiv.org/abs/1807.06211} {arXiv:1807.06211 [astro-ph.CO]}
  \BibitemShut {NoStop}%
\bibitem [{\citenamefont {Kofman}\ \emph {et~al.}(1997)\citenamefont {Kofman},
  \citenamefont {Linde},\ and\ \citenamefont {Starobinsky}}]{Kofman:1997yn}%
  \BibitemOpen
  \bibfield  {author} {\bibinfo {author} {\bibfnamefont {L.}~\bibnamefont
  {Kofman}}, \bibinfo {author} {\bibfnamefont {A.~D.}\ \bibnamefont {Linde}}, \
  and\ \bibinfo {author} {\bibfnamefont {A.~A.}\ \bibnamefont {Starobinsky}},\
  }\href {\doibase 10.1103/PhysRevD.56.3258} {\bibfield  {journal} {\bibinfo
  {journal} {Phys. Rev. D}\ }\textbf {\bibinfo {volume} {56}},\ \bibinfo
  {pages} {3258} (\bibinfo {year} {1997})},\ \Eprint
  {http://arxiv.org/abs/hep-ph/9704452} {arXiv:hep-ph/9704452} \BibitemShut
  {NoStop}%
\bibitem [{\citenamefont {Allahverdi}\ \emph {et~al.}(2010)\citenamefont
  {Allahverdi}, \citenamefont {Brandenberger}, \citenamefont {Cyr-Racine},\
  and\ \citenamefont {Mazumdar}}]{Allahverdi:2010xz}%
  \BibitemOpen
  \bibfield  {author} {\bibinfo {author} {\bibfnamefont {R.}~\bibnamefont
  {Allahverdi}}, \bibinfo {author} {\bibfnamefont {R.}~\bibnamefont
  {Brandenberger}}, \bibinfo {author} {\bibfnamefont {F.-Y.}\ \bibnamefont
  {Cyr-Racine}}, \ and\ \bibinfo {author} {\bibfnamefont {A.}~\bibnamefont
  {Mazumdar}},\ }\href {\doibase 10.1146/annurev.nucl.012809.104511} {\bibfield
   {journal} {\bibinfo  {journal} {Ann. Rev. Nucl. Part. Sci.}\ }\textbf
  {\bibinfo {volume} {60}},\ \bibinfo {pages} {27} (\bibinfo {year} {2010})},\
  \Eprint {http://arxiv.org/abs/1001.2600} {arXiv:1001.2600 [hep-th]}
  \BibitemShut {NoStop}%
\bibitem [{\citenamefont {Lozanov}(2019)}]{Lozanov:2019jxc}%
  \BibitemOpen
  \bibfield  {author} {\bibinfo {author} {\bibfnamefont {K.~D.}\ \bibnamefont
  {Lozanov}},\ }\href@noop {} {\  (\bibinfo {year} {2019})},\ \Eprint
  {http://arxiv.org/abs/1907.04402} {arXiv:1907.04402 [astro-ph.CO]}
  \BibitemShut {NoStop}%
\bibitem [{\citenamefont {Bassett}\ \emph {et~al.}(2006)\citenamefont
  {Bassett}, \citenamefont {Tsujikawa},\ and\ \citenamefont
  {Wands}}]{Bassett:2005xm}%
  \BibitemOpen
  \bibfield  {author} {\bibinfo {author} {\bibfnamefont {B.~A.}\ \bibnamefont
  {Bassett}}, \bibinfo {author} {\bibfnamefont {S.}~\bibnamefont {Tsujikawa}},
  \ and\ \bibinfo {author} {\bibfnamefont {D.}~\bibnamefont {Wands}},\ }\href
  {\doibase 10.1103/RevModPhys.78.537} {\bibfield  {journal} {\bibinfo
  {journal} {Rev. Mod. Phys.}\ }\textbf {\bibinfo {volume} {78}},\ \bibinfo
  {pages} {537} (\bibinfo {year} {2006})},\ \Eprint
  {http://arxiv.org/abs/astro-ph/0507632} {arXiv:astro-ph/0507632} \BibitemShut
  {NoStop}%
\bibitem [{\citenamefont {Amin}\ \emph {et~al.}(2014)\citenamefont {Amin},
  \citenamefont {Hertzberg}, \citenamefont {Kaiser},\ and\ \citenamefont
  {Karouby}}]{Amin:2014eta}%
  \BibitemOpen
  \bibfield  {author} {\bibinfo {author} {\bibfnamefont {M.~A.}\ \bibnamefont
  {Amin}}, \bibinfo {author} {\bibfnamefont {M.~P.}\ \bibnamefont {Hertzberg}},
  \bibinfo {author} {\bibfnamefont {D.~I.}\ \bibnamefont {Kaiser}}, \ and\
  \bibinfo {author} {\bibfnamefont {J.}~\bibnamefont {Karouby}},\ }\href
  {\doibase 10.1142/S0218271815300037} {\bibfield  {journal} {\bibinfo
  {journal} {Int. J. Mod. Phys. D}\ }\textbf {\bibinfo {volume} {24}},\
  \bibinfo {pages} {1530003} (\bibinfo {year} {2014})},\ \Eprint
  {http://arxiv.org/abs/1410.3808} {arXiv:1410.3808 [hep-ph]} \BibitemShut
  {NoStop}%
\bibitem [{\citenamefont {Shandera}\ \emph {et~al.}(2019)\citenamefont
  {Shandera} \emph {et~al.}}]{Shandera:2019ufi}%
  \BibitemOpen
  \bibfield  {author} {\bibinfo {author} {\bibfnamefont {S.}~\bibnamefont
  {Shandera}} \emph {et~al.},\ }\href@noop {} {\bibfield  {journal} {\bibinfo
  {journal} {Bull. Am. Astron. Soc.}\ }\textbf {\bibinfo {volume} {51}},\
  \bibinfo {pages} {338} (\bibinfo {year} {2019})},\ \Eprint
  {http://arxiv.org/abs/1903.04700} {arXiv:1903.04700 [astro-ph.CO]}
  \BibitemShut {NoStop}%
\bibitem [{\citenamefont {Bartolo}\ \emph {et~al.}(2022)\citenamefont {Bartolo}
  \emph {et~al.}}]{LISACosmologyWorkingGroup:2022kbp}%
  \BibitemOpen
  \bibfield  {author} {\bibinfo {author} {\bibfnamefont {N.}~\bibnamefont
  {Bartolo}} \emph {et~al.} (\bibinfo {collaboration} {LISA Cosmology Working
  Group}),\ }\href@noop {} {\  (\bibinfo {year} {2022})},\ \Eprint
  {http://arxiv.org/abs/2201.08782} {arXiv:2201.08782 [astro-ph.CO]}
  \BibitemShut {NoStop}%
\bibitem [{\citenamefont {Ananda}\ \emph {et~al.}(2007)\citenamefont {Ananda},
  \citenamefont {Clarkson},\ and\ \citenamefont {Wands}}]{Ananda:2006af}%
  \BibitemOpen
  \bibfield  {author} {\bibinfo {author} {\bibfnamefont {K.~N.}\ \bibnamefont
  {Ananda}}, \bibinfo {author} {\bibfnamefont {C.}~\bibnamefont {Clarkson}}, \
  and\ \bibinfo {author} {\bibfnamefont {D.}~\bibnamefont {Wands}},\ }\href
  {\doibase 10.1103/PhysRevD.75.123518} {\bibfield  {journal} {\bibinfo
  {journal} {Phys. Rev. D}\ }\textbf {\bibinfo {volume} {75}},\ \bibinfo
  {pages} {123518} (\bibinfo {year} {2007})},\ \Eprint
  {http://arxiv.org/abs/gr-qc/0612013} {arXiv:gr-qc/0612013} \BibitemShut
  {NoStop}%
\bibitem [{\citenamefont {Baumann}\ \emph {et~al.}(2007)\citenamefont
  {Baumann}, \citenamefont {Steinhardt}, \citenamefont {Takahashi},\ and\
  \citenamefont {Ichiki}}]{Baumann:2007zm}%
  \BibitemOpen
  \bibfield  {author} {\bibinfo {author} {\bibfnamefont {D.}~\bibnamefont
  {Baumann}}, \bibinfo {author} {\bibfnamefont {P.~J.}\ \bibnamefont
  {Steinhardt}}, \bibinfo {author} {\bibfnamefont {K.}~\bibnamefont
  {Takahashi}}, \ and\ \bibinfo {author} {\bibfnamefont {K.}~\bibnamefont
  {Ichiki}},\ }\href {\doibase 10.1103/PhysRevD.76.084019} {\bibfield
  {journal} {\bibinfo  {journal} {Phys. Rev. D}\ }\textbf {\bibinfo {volume}
  {76}},\ \bibinfo {pages} {084019} (\bibinfo {year} {2007})},\ \Eprint
  {http://arxiv.org/abs/hep-th/0703290} {arXiv:hep-th/0703290} \BibitemShut
  {NoStop}%
\bibitem [{\citenamefont {Saito}\ and\ \citenamefont
  {Yokoyama}(2009)}]{Saito:2008jc}%
  \BibitemOpen
  \bibfield  {author} {\bibinfo {author} {\bibfnamefont {R.}~\bibnamefont
  {Saito}}\ and\ \bibinfo {author} {\bibfnamefont {J.}~\bibnamefont
  {Yokoyama}},\ }\href {\doibase 10.1103/PhysRevLett.102.161101} {\bibfield
  {journal} {\bibinfo  {journal} {Phys. Rev. Lett.}\ }\textbf {\bibinfo
  {volume} {102}},\ \bibinfo {pages} {161101} (\bibinfo {year} {2009})},\
  \bibinfo {note} {[Erratum: Phys.Rev.Lett. 107, 069901 (2011)]},\ \Eprint
  {http://arxiv.org/abs/0812.4339} {arXiv:0812.4339 [astro-ph]} \BibitemShut
  {NoStop}%
\bibitem [{\citenamefont {Inomata}\ \emph
  {et~al.}(2019{\natexlab{a}})\citenamefont {Inomata}, \citenamefont {Kohri},
  \citenamefont {Nakama},\ and\ \citenamefont {Terada}}]{Inomata:2019ivs}%
  \BibitemOpen
  \bibfield  {author} {\bibinfo {author} {\bibfnamefont {K.}~\bibnamefont
  {Inomata}}, \bibinfo {author} {\bibfnamefont {K.}~\bibnamefont {Kohri}},
  \bibinfo {author} {\bibfnamefont {T.}~\bibnamefont {Nakama}}, \ and\ \bibinfo
  {author} {\bibfnamefont {T.}~\bibnamefont {Terada}},\ }\href {\doibase
  10.1103/PhysRevD.100.043532} {\bibfield  {journal} {\bibinfo  {journal}
  {Phys. Rev. D}\ }\textbf {\bibinfo {volume} {100}},\ \bibinfo {pages}
  {043532} (\bibinfo {year} {2019}{\natexlab{a}})},\ \Eprint
  {http://arxiv.org/abs/1904.12879} {arXiv:1904.12879 [astro-ph.CO]}
  \BibitemShut {NoStop}%
\bibitem [{\citenamefont {Inomata}\ \emph
  {et~al.}(2019{\natexlab{b}})\citenamefont {Inomata}, \citenamefont {Kohri},
  \citenamefont {Nakama},\ and\ \citenamefont {Terada}}]{Inomata:2019zqy}%
  \BibitemOpen
  \bibfield  {author} {\bibinfo {author} {\bibfnamefont {K.}~\bibnamefont
  {Inomata}}, \bibinfo {author} {\bibfnamefont {K.}~\bibnamefont {Kohri}},
  \bibinfo {author} {\bibfnamefont {T.}~\bibnamefont {Nakama}}, \ and\ \bibinfo
  {author} {\bibfnamefont {T.}~\bibnamefont {Terada}},\ }\href {\doibase
  10.1088/1475-7516/2019/10/071} {\bibfield  {journal} {\bibinfo  {journal}
  {JCAP}\ }\textbf {\bibinfo {volume} {10}},\ \bibinfo {pages} {071} (\bibinfo
  {year} {2019}{\natexlab{b}})},\ \Eprint {http://arxiv.org/abs/1904.12878}
  {arXiv:1904.12878 [astro-ph.CO]} \BibitemShut {NoStop}%
\bibitem [{\citenamefont {Inomata}\ \emph {et~al.}(2021)\citenamefont
  {Inomata}, \citenamefont {Kawasaki}, \citenamefont {Mukaida},\ and\
  \citenamefont {Yanagida}}]{Inomata:2020xad}%
  \BibitemOpen
  \bibfield  {author} {\bibinfo {author} {\bibfnamefont {K.}~\bibnamefont
  {Inomata}}, \bibinfo {author} {\bibfnamefont {M.}~\bibnamefont {Kawasaki}},
  \bibinfo {author} {\bibfnamefont {K.}~\bibnamefont {Mukaida}}, \ and\
  \bibinfo {author} {\bibfnamefont {T.~T.}\ \bibnamefont {Yanagida}},\ }\href
  {\doibase 10.1103/PhysRevLett.126.131301} {\bibfield  {journal} {\bibinfo
  {journal} {Phys. Rev. Lett.}\ }\textbf {\bibinfo {volume} {126}},\ \bibinfo
  {pages} {131301} (\bibinfo {year} {2021})},\ \Eprint
  {http://arxiv.org/abs/2011.01270} {arXiv:2011.01270 [astro-ph.CO]}
  \BibitemShut {NoStop}%
\bibitem [{\citenamefont {Dom\`enech}\ \emph {et~al.}(2020)\citenamefont
  {Dom\`enech}, \citenamefont {Pi},\ and\ \citenamefont
  {Sasaki}}]{Domenech:2020kqm}%
  \BibitemOpen
  \bibfield  {author} {\bibinfo {author} {\bibfnamefont {G.}~\bibnamefont
  {Dom\`enech}}, \bibinfo {author} {\bibfnamefont {S.}~\bibnamefont {Pi}}, \
  and\ \bibinfo {author} {\bibfnamefont {M.}~\bibnamefont {Sasaki}},\ }\href
  {\doibase 10.1088/1475-7516/2020/08/017} {\bibfield  {journal} {\bibinfo
  {journal} {JCAP}\ }\textbf {\bibinfo {volume} {08}},\ \bibinfo {pages} {017}
  (\bibinfo {year} {2020})},\ \Eprint {http://arxiv.org/abs/2005.12314}
  {arXiv:2005.12314 [gr-qc]} \BibitemShut {NoStop}%
\bibitem [{\citenamefont {Sugiyama}\ \emph {et~al.}(2021)\citenamefont
  {Sugiyama}, \citenamefont {Takhistov}, \citenamefont {Vitagliano},
  \citenamefont {Kusenko}, \citenamefont {Sasaki},\ and\ \citenamefont
  {Takada}}]{Sugiyama:2020roc}%
  \BibitemOpen
  \bibfield  {author} {\bibinfo {author} {\bibfnamefont {S.}~\bibnamefont
  {Sugiyama}}, \bibinfo {author} {\bibfnamefont {V.}~\bibnamefont {Takhistov}},
  \bibinfo {author} {\bibfnamefont {E.}~\bibnamefont {Vitagliano}}, \bibinfo
  {author} {\bibfnamefont {A.}~\bibnamefont {Kusenko}}, \bibinfo {author}
  {\bibfnamefont {M.}~\bibnamefont {Sasaki}}, \ and\ \bibinfo {author}
  {\bibfnamefont {M.}~\bibnamefont {Takada}},\ }\href {\doibase
  10.1016/j.physletb.2021.136097} {\bibfield  {journal} {\bibinfo  {journal}
  {Phys. Lett. B}\ }\textbf {\bibinfo {volume} {814}},\ \bibinfo {pages}
  {136097} (\bibinfo {year} {2021})},\ \Eprint
  {http://arxiv.org/abs/2010.02189} {arXiv:2010.02189 [astro-ph.CO]}
  \BibitemShut {NoStop}%
\bibitem [{\citenamefont {Adshead}\ \emph {et~al.}(2021)\citenamefont
  {Adshead}, \citenamefont {Lozanov},\ and\ \citenamefont
  {Weiner}}]{Adshead:2021hnm}%
  \BibitemOpen
  \bibfield  {author} {\bibinfo {author} {\bibfnamefont {P.}~\bibnamefont
  {Adshead}}, \bibinfo {author} {\bibfnamefont {K.~D.}\ \bibnamefont
  {Lozanov}}, \ and\ \bibinfo {author} {\bibfnamefont {Z.~J.}\ \bibnamefont
  {Weiner}},\ }\href {\doibase 10.1088/1475-7516/2021/10/080} {\bibfield
  {journal} {\bibinfo  {journal} {JCAP}\ }\textbf {\bibinfo {volume} {10}},\
  \bibinfo {pages} {080} (\bibinfo {year} {2021})},\ \Eprint
  {http://arxiv.org/abs/2105.01659} {arXiv:2105.01659 [astro-ph.CO]}
  \BibitemShut {NoStop}%
\bibitem [{\citenamefont {Dom\`enech}\ \emph
  {et~al.}(2021{\natexlab{a}})\citenamefont {Dom\`enech}, \citenamefont
  {Takhistov},\ and\ \citenamefont {Sasaki}}]{Domenech:2021wkk}%
  \BibitemOpen
  \bibfield  {author} {\bibinfo {author} {\bibfnamefont {G.}~\bibnamefont
  {Dom\`enech}}, \bibinfo {author} {\bibfnamefont {V.}~\bibnamefont
  {Takhistov}}, \ and\ \bibinfo {author} {\bibfnamefont {M.}~\bibnamefont
  {Sasaki}},\ }\href {\doibase 10.1016/j.physletb.2021.136722} {\bibfield
  {journal} {\bibinfo  {journal} {Phys. Lett. B}\ }\textbf {\bibinfo {volume}
  {823}},\ \bibinfo {pages} {136722} (\bibinfo {year} {2021}{\natexlab{a}})},\
  \Eprint {http://arxiv.org/abs/2105.06816} {arXiv:2105.06816 [astro-ph.CO]}
  \BibitemShut {NoStop}%
\bibitem [{\citenamefont {Dom\`enech}(2021)}]{Domenech:2021ztg}%
  \BibitemOpen
  \bibfield  {author} {\bibinfo {author} {\bibfnamefont {G.}~\bibnamefont
  {Dom\`enech}},\ }\href {\doibase 10.3390/universe7110398} {\bibfield
  {journal} {\bibinfo  {journal} {Universe}\ }\textbf {\bibinfo {volume} {7}},\
  \bibinfo {pages} {398} (\bibinfo {year} {2021})},\ \Eprint
  {http://arxiv.org/abs/2109.01398} {arXiv:2109.01398 [gr-qc]} \BibitemShut
  {NoStop}%
\bibitem [{\citenamefont {Nomura}\ \emph {et~al.}(2018)\citenamefont {Nomura},
  \citenamefont {Watari},\ and\ \citenamefont {Yamazaki}}]{Nomura:2017ehb}%
  \BibitemOpen
  \bibfield  {author} {\bibinfo {author} {\bibfnamefont {Y.}~\bibnamefont
  {Nomura}}, \bibinfo {author} {\bibfnamefont {T.}~\bibnamefont {Watari}}, \
  and\ \bibinfo {author} {\bibfnamefont {M.}~\bibnamefont {Yamazaki}},\ }\href
  {\doibase 10.1016/j.physletb.2017.11.052} {\bibfield  {journal} {\bibinfo
  {journal} {Phys. Lett. B}\ }\textbf {\bibinfo {volume} {776}},\ \bibinfo
  {pages} {227} (\bibinfo {year} {2018})},\ \Eprint
  {http://arxiv.org/abs/1706.08522} {arXiv:1706.08522 [hep-ph]} \BibitemShut
  {NoStop}%
\bibitem [{\citenamefont {Nomura}\ and\ \citenamefont
  {Yamazaki}(2018)}]{Nomura:2017zqj}%
  \BibitemOpen
  \bibfield  {author} {\bibinfo {author} {\bibfnamefont {Y.}~\bibnamefont
  {Nomura}}\ and\ \bibinfo {author} {\bibfnamefont {M.}~\bibnamefont
  {Yamazaki}},\ }\href {\doibase 10.1016/j.physletb.2018.02.071} {\bibfield
  {journal} {\bibinfo  {journal} {Phys. Lett. B}\ }\textbf {\bibinfo {volume}
  {780}},\ \bibinfo {pages} {106} (\bibinfo {year} {2018})},\ \Eprint
  {http://arxiv.org/abs/1711.10490} {arXiv:1711.10490 [hep-ph]} \BibitemShut
  {NoStop}%
\bibitem [{\citenamefont {Silverstein}\ and\ \citenamefont
  {Westphal}(2008)}]{Silverstein:2008sg}%
  \BibitemOpen
  \bibfield  {author} {\bibinfo {author} {\bibfnamefont {E.}~\bibnamefont
  {Silverstein}}\ and\ \bibinfo {author} {\bibfnamefont {A.}~\bibnamefont
  {Westphal}},\ }\href {\doibase 10.1103/PhysRevD.78.106003} {\bibfield
  {journal} {\bibinfo  {journal} {Phys. Rev. D}\ }\textbf {\bibinfo {volume}
  {78}},\ \bibinfo {pages} {106003} (\bibinfo {year} {2008})},\ \Eprint
  {http://arxiv.org/abs/0803.3085} {arXiv:0803.3085 [hep-th]} \BibitemShut
  {NoStop}%
\bibitem [{\citenamefont {McAllister}\ \emph {et~al.}(2014)\citenamefont
  {McAllister}, \citenamefont {Silverstein}, \citenamefont {Westphal},\ and\
  \citenamefont {Wrase}}]{McAllister:2014mpa}%
  \BibitemOpen
  \bibfield  {author} {\bibinfo {author} {\bibfnamefont {L.}~\bibnamefont
  {McAllister}}, \bibinfo {author} {\bibfnamefont {E.}~\bibnamefont
  {Silverstein}}, \bibinfo {author} {\bibfnamefont {A.}~\bibnamefont
  {Westphal}}, \ and\ \bibinfo {author} {\bibfnamefont {T.}~\bibnamefont
  {Wrase}},\ }\href {\doibase 10.1007/JHEP09(2014)123} {\bibfield  {journal}
  {\bibinfo  {journal} {JHEP}\ }\textbf {\bibinfo {volume} {09}},\ \bibinfo
  {pages} {123} (\bibinfo {year} {2014})},\ \Eprint
  {http://arxiv.org/abs/1405.3652} {arXiv:1405.3652 [hep-th]} \BibitemShut
  {NoStop}%
\bibitem [{\citenamefont {Adams}\ \emph {et~al.}(1993)\citenamefont {Adams},
  \citenamefont {Bond}, \citenamefont {Freese}, \citenamefont {Frieman},\ and\
  \citenamefont {Olinto}}]{Adams:1992bn}%
  \BibitemOpen
  \bibfield  {author} {\bibinfo {author} {\bibfnamefont {F.~C.}\ \bibnamefont
  {Adams}}, \bibinfo {author} {\bibfnamefont {J.~R.}\ \bibnamefont {Bond}},
  \bibinfo {author} {\bibfnamefont {K.}~\bibnamefont {Freese}}, \bibinfo
  {author} {\bibfnamefont {J.~A.}\ \bibnamefont {Frieman}}, \ and\ \bibinfo
  {author} {\bibfnamefont {A.~V.}\ \bibnamefont {Olinto}},\ }\href {\doibase
  10.1103/PhysRevD.47.426} {\bibfield  {journal} {\bibinfo  {journal} {Phys.
  Rev. D}\ }\textbf {\bibinfo {volume} {47}},\ \bibinfo {pages} {426} (\bibinfo
  {year} {1993})},\ \Eprint {http://arxiv.org/abs/hep-ph/9207245}
  {arXiv:hep-ph/9207245} \BibitemShut {NoStop}%
\bibitem [{\citenamefont {Freese}\ \emph {et~al.}(1990)\citenamefont {Freese},
  \citenamefont {Frieman},\ and\ \citenamefont {Olinto}}]{Freese:1990rb}%
  \BibitemOpen
  \bibfield  {author} {\bibinfo {author} {\bibfnamefont {K.}~\bibnamefont
  {Freese}}, \bibinfo {author} {\bibfnamefont {J.~A.}\ \bibnamefont {Frieman}},
  \ and\ \bibinfo {author} {\bibfnamefont {A.~V.}\ \bibnamefont {Olinto}},\
  }\href {\doibase 10.1103/PhysRevLett.65.3233} {\bibfield  {journal} {\bibinfo
   {journal} {Phys. Rev. Lett.}\ }\textbf {\bibinfo {volume} {65}},\ \bibinfo
  {pages} {3233} (\bibinfo {year} {1990})}\BibitemShut {NoStop}%
\bibitem [{\citenamefont {Antusch}\ \emph {et~al.}(2017)\citenamefont
  {Antusch}, \citenamefont {Cefala},\ and\ \citenamefont
  {Orani}}]{Antusch:2016con}%
  \BibitemOpen
  \bibfield  {author} {\bibinfo {author} {\bibfnamefont {S.}~\bibnamefont
  {Antusch}}, \bibinfo {author} {\bibfnamefont {F.}~\bibnamefont {Cefala}}, \
  and\ \bibinfo {author} {\bibfnamefont {S.}~\bibnamefont {Orani}},\ }\href
  {\doibase 10.1103/PhysRevLett.118.011303} {\bibfield  {journal} {\bibinfo
  {journal} {Phys. Rev. Lett.}\ }\textbf {\bibinfo {volume} {118}},\ \bibinfo
  {pages} {011303} (\bibinfo {year} {2017})},\ \bibinfo {note} {[Erratum:
  Phys.Rev.Lett. 120, 219901 (2018)]},\ \Eprint
  {http://arxiv.org/abs/1607.01314} {arXiv:1607.01314 [astro-ph.CO]}
  \BibitemShut {NoStop}%
\bibitem [{\citenamefont {Antusch}\ \emph {et~al.}(2018)\citenamefont
  {Antusch}, \citenamefont {Cefala}, \citenamefont {Krippendorf}, \citenamefont
  {Muia}, \citenamefont {Orani},\ and\ \citenamefont
  {Quevedo}}]{Antusch:2017flz}%
  \BibitemOpen
  \bibfield  {author} {\bibinfo {author} {\bibfnamefont {S.}~\bibnamefont
  {Antusch}}, \bibinfo {author} {\bibfnamefont {F.}~\bibnamefont {Cefala}},
  \bibinfo {author} {\bibfnamefont {S.}~\bibnamefont {Krippendorf}}, \bibinfo
  {author} {\bibfnamefont {F.}~\bibnamefont {Muia}}, \bibinfo {author}
  {\bibfnamefont {S.}~\bibnamefont {Orani}}, \ and\ \bibinfo {author}
  {\bibfnamefont {F.}~\bibnamefont {Quevedo}},\ }\href {\doibase
  10.1007/JHEP01(2018)083} {\bibfield  {journal} {\bibinfo  {journal} {JHEP}\
  }\textbf {\bibinfo {volume} {01}},\ \bibinfo {pages} {083} (\bibinfo {year}
  {2018})},\ \Eprint {http://arxiv.org/abs/1708.08922} {arXiv:1708.08922
  [hep-th]} \BibitemShut {NoStop}%
\bibitem [{\citenamefont {Bogolyubsky}\ and\ \citenamefont
  {Makhankov}(1976)}]{Bogolyubsky:1976nx}%
  \BibitemOpen
  \bibfield  {author} {\bibinfo {author} {\bibfnamefont {I.~L.}\ \bibnamefont
  {Bogolyubsky}}\ and\ \bibinfo {author} {\bibfnamefont {V.~G.}\ \bibnamefont
  {Makhankov}},\ }\href@noop {} {\bibfield  {journal} {\bibinfo  {journal}
  {JETP Lett.}\ }\textbf {\bibinfo {volume} {24}},\ \bibinfo {pages} {12}
  (\bibinfo {year} {1976})}\BibitemShut {NoStop}%
\bibitem [{\citenamefont {Gleiser}(1994)}]{Gleiser:1993pt}%
  \BibitemOpen
  \bibfield  {author} {\bibinfo {author} {\bibfnamefont {M.}~\bibnamefont
  {Gleiser}},\ }\href {\doibase 10.1103/PhysRevD.49.2978} {\bibfield  {journal}
  {\bibinfo  {journal} {Phys. Rev. D}\ }\textbf {\bibinfo {volume} {49}},\
  \bibinfo {pages} {2978} (\bibinfo {year} {1994})},\ \Eprint
  {http://arxiv.org/abs/hep-ph/9308279} {arXiv:hep-ph/9308279} \BibitemShut
  {NoStop}%
\bibitem [{\citenamefont {Kasuya}\ \emph {et~al.}(2003)\citenamefont {Kasuya},
  \citenamefont {Kawasaki},\ and\ \citenamefont {Takahashi}}]{Kasuya:2002zs}%
  \BibitemOpen
  \bibfield  {author} {\bibinfo {author} {\bibfnamefont {S.}~\bibnamefont
  {Kasuya}}, \bibinfo {author} {\bibfnamefont {M.}~\bibnamefont {Kawasaki}}, \
  and\ \bibinfo {author} {\bibfnamefont {F.}~\bibnamefont {Takahashi}},\ }\href
  {\doibase 10.1016/S0370-2693(03)00344-7} {\bibfield  {journal} {\bibinfo
  {journal} {Phys. Lett. B}\ }\textbf {\bibinfo {volume} {559}},\ \bibinfo
  {pages} {99} (\bibinfo {year} {2003})},\ \Eprint
  {http://arxiv.org/abs/hep-ph/0209358} {arXiv:hep-ph/0209358} \BibitemShut
  {NoStop}%
\bibitem [{\citenamefont {Copeland}\ \emph {et~al.}(1995)\citenamefont
  {Copeland}, \citenamefont {Gleiser},\ and\ \citenamefont
  {Muller}}]{Copeland:1995fq}%
  \BibitemOpen
  \bibfield  {author} {\bibinfo {author} {\bibfnamefont {E.~J.}\ \bibnamefont
  {Copeland}}, \bibinfo {author} {\bibfnamefont {M.}~\bibnamefont {Gleiser}}, \
  and\ \bibinfo {author} {\bibfnamefont {H.~R.}\ \bibnamefont {Muller}},\
  }\href {\doibase 10.1103/PhysRevD.52.1920} {\bibfield  {journal} {\bibinfo
  {journal} {Phys. Rev. D}\ }\textbf {\bibinfo {volume} {52}},\ \bibinfo
  {pages} {1920} (\bibinfo {year} {1995})},\ \Eprint
  {http://arxiv.org/abs/hep-ph/9503217} {arXiv:hep-ph/9503217} \BibitemShut
  {NoStop}%
\bibitem [{\citenamefont {Broadhead}\ and\ \citenamefont
  {McDonald}(2005)}]{Broadhead:2005hn}%
  \BibitemOpen
  \bibfield  {author} {\bibinfo {author} {\bibfnamefont {M.}~\bibnamefont
  {Broadhead}}\ and\ \bibinfo {author} {\bibfnamefont {J.}~\bibnamefont
  {McDonald}},\ }\href {\doibase 10.1103/PhysRevD.72.043519} {\bibfield
  {journal} {\bibinfo  {journal} {Phys. Rev. D}\ }\textbf {\bibinfo {volume}
  {72}},\ \bibinfo {pages} {043519} (\bibinfo {year} {2005})},\ \Eprint
  {http://arxiv.org/abs/hep-ph/0503081} {arXiv:hep-ph/0503081} \BibitemShut
  {NoStop}%
\bibitem [{\citenamefont {Amin}(2010)}]{Amin:2010xe}%
  \BibitemOpen
  \bibfield  {author} {\bibinfo {author} {\bibfnamefont {M.~A.}\ \bibnamefont
  {Amin}},\ }\href@noop {} {\  (\bibinfo {year} {2010})},\ \Eprint
  {http://arxiv.org/abs/1006.3075} {arXiv:1006.3075 [astro-ph.CO]} \BibitemShut
  {NoStop}%
\bibitem [{\citenamefont {Amin}\ \emph {et~al.}(2010)\citenamefont {Amin},
  \citenamefont {Easther},\ and\ \citenamefont {Finkel}}]{Amin:2010dc}%
  \BibitemOpen
  \bibfield  {author} {\bibinfo {author} {\bibfnamefont {M.~A.}\ \bibnamefont
  {Amin}}, \bibinfo {author} {\bibfnamefont {R.}~\bibnamefont {Easther}}, \
  and\ \bibinfo {author} {\bibfnamefont {H.}~\bibnamefont {Finkel}},\ }\href
  {\doibase 10.1088/1475-7516/2010/12/001} {\bibfield  {journal} {\bibinfo
  {journal} {JCAP}\ }\textbf {\bibinfo {volume} {12}},\ \bibinfo {pages} {001}
  (\bibinfo {year} {2010})},\ \Eprint {http://arxiv.org/abs/1009.2505}
  {arXiv:1009.2505 [astro-ph.CO]} \BibitemShut {NoStop}%
\bibitem [{\citenamefont {Amin}\ \emph {et~al.}(2012)\citenamefont {Amin},
  \citenamefont {Easther}, \citenamefont {Finkel}, \citenamefont {Flauger},\
  and\ \citenamefont {Hertzberg}}]{Amin:2011hj}%
  \BibitemOpen
  \bibfield  {author} {\bibinfo {author} {\bibfnamefont {M.~A.}\ \bibnamefont
  {Amin}}, \bibinfo {author} {\bibfnamefont {R.}~\bibnamefont {Easther}},
  \bibinfo {author} {\bibfnamefont {H.}~\bibnamefont {Finkel}}, \bibinfo
  {author} {\bibfnamefont {R.}~\bibnamefont {Flauger}}, \ and\ \bibinfo
  {author} {\bibfnamefont {M.~P.}\ \bibnamefont {Hertzberg}},\ }\href {\doibase
  10.1103/PhysRevLett.108.241302} {\bibfield  {journal} {\bibinfo  {journal}
  {Phys. Rev. Lett.}\ }\textbf {\bibinfo {volume} {108}},\ \bibinfo {pages}
  {241302} (\bibinfo {year} {2012})},\ \Eprint {http://arxiv.org/abs/1106.3335}
  {arXiv:1106.3335 [astro-ph.CO]} \BibitemShut {NoStop}%
\bibitem [{\citenamefont {Lozanov}\ and\ \citenamefont
  {Amin}(2019)}]{Lozanov:2019ylm}%
  \BibitemOpen
  \bibfield  {author} {\bibinfo {author} {\bibfnamefont {K.~D.}\ \bibnamefont
  {Lozanov}}\ and\ \bibinfo {author} {\bibfnamefont {M.~A.}\ \bibnamefont
  {Amin}},\ }\href {\doibase 10.1103/PhysRevD.99.123504} {\bibfield  {journal}
  {\bibinfo  {journal} {Phys. Rev. D}\ }\textbf {\bibinfo {volume} {99}},\
  \bibinfo {pages} {123504} (\bibinfo {year} {2019})},\ \Eprint
  {http://arxiv.org/abs/1902.06736} {arXiv:1902.06736 [astro-ph.CO]}
  \BibitemShut {NoStop}%
\bibitem [{\citenamefont {Zhou}\ \emph {et~al.}(2013)\citenamefont {Zhou},
  \citenamefont {Copeland}, \citenamefont {Easther}, \citenamefont {Finkel},
  \citenamefont {Mou},\ and\ \citenamefont {Saffin}}]{Zhou:2013tsa}%
  \BibitemOpen
  \bibfield  {author} {\bibinfo {author} {\bibfnamefont {S.-Y.}\ \bibnamefont
  {Zhou}}, \bibinfo {author} {\bibfnamefont {E.~J.}\ \bibnamefont {Copeland}},
  \bibinfo {author} {\bibfnamefont {R.}~\bibnamefont {Easther}}, \bibinfo
  {author} {\bibfnamefont {H.}~\bibnamefont {Finkel}}, \bibinfo {author}
  {\bibfnamefont {Z.-G.}\ \bibnamefont {Mou}}, \ and\ \bibinfo {author}
  {\bibfnamefont {P.~M.}\ \bibnamefont {Saffin}},\ }\href {\doibase
  10.1007/JHEP10(2013)026} {\bibfield  {journal} {\bibinfo  {journal} {JHEP}\
  }\textbf {\bibinfo {volume} {10}},\ \bibinfo {pages} {026} (\bibinfo {year}
  {2013})},\ \Eprint {http://arxiv.org/abs/1304.6094} {arXiv:1304.6094
  [astro-ph.CO]} \BibitemShut {NoStop}%
\bibitem [{\citenamefont {Liu}\ \emph {et~al.}(2018)\citenamefont {Liu},
  \citenamefont {Guo}, \citenamefont {Cai},\ and\ \citenamefont
  {Shiu}}]{Liu:2017hua}%
  \BibitemOpen
  \bibfield  {author} {\bibinfo {author} {\bibfnamefont {J.}~\bibnamefont
  {Liu}}, \bibinfo {author} {\bibfnamefont {Z.-K.}\ \bibnamefont {Guo}},
  \bibinfo {author} {\bibfnamefont {R.-G.}\ \bibnamefont {Cai}}, \ and\
  \bibinfo {author} {\bibfnamefont {G.}~\bibnamefont {Shiu}},\ }\href {\doibase
  10.1103/PhysRevLett.120.031301} {\bibfield  {journal} {\bibinfo  {journal}
  {Phys. Rev. Lett.}\ }\textbf {\bibinfo {volume} {120}},\ \bibinfo {pages}
  {031301} (\bibinfo {year} {2018})},\ \Eprint
  {http://arxiv.org/abs/1707.09841} {arXiv:1707.09841 [astro-ph.CO]}
  \BibitemShut {NoStop}%
\bibitem [{\citenamefont {Amin}\ \emph {et~al.}(2018)\citenamefont {Amin},
  \citenamefont {Braden}, \citenamefont {Copeland}, \citenamefont {Giblin},
  \citenamefont {Solorio}, \citenamefont {Weiner},\ and\ \citenamefont
  {Zhou}}]{Amin:2018xfe}%
  \BibitemOpen
  \bibfield  {author} {\bibinfo {author} {\bibfnamefont {M.~A.}\ \bibnamefont
  {Amin}}, \bibinfo {author} {\bibfnamefont {J.}~\bibnamefont {Braden}},
  \bibinfo {author} {\bibfnamefont {E.~J.}\ \bibnamefont {Copeland}}, \bibinfo
  {author} {\bibfnamefont {J.~T.}\ \bibnamefont {Giblin}}, \bibinfo {author}
  {\bibfnamefont {C.}~\bibnamefont {Solorio}}, \bibinfo {author} {\bibfnamefont
  {Z.~J.}\ \bibnamefont {Weiner}}, \ and\ \bibinfo {author} {\bibfnamefont
  {S.-Y.}\ \bibnamefont {Zhou}},\ }\href {\doibase 10.1103/PhysRevD.98.024040}
  {\bibfield  {journal} {\bibinfo  {journal} {Phys. Rev. D}\ }\textbf {\bibinfo
  {volume} {98}},\ \bibinfo {pages} {024040} (\bibinfo {year} {2018})},\
  \Eprint {http://arxiv.org/abs/1803.08047} {arXiv:1803.08047 [astro-ph.CO]}
  \BibitemShut {NoStop}%
\bibitem [{\citenamefont {Hiramatsu}\ \emph {et~al.}(2021)\citenamefont
  {Hiramatsu}, \citenamefont {Sfakianakis},\ and\ \citenamefont
  {Yamaguchi}}]{Hiramatsu:2020obh}%
  \BibitemOpen
  \bibfield  {author} {\bibinfo {author} {\bibfnamefont {T.}~\bibnamefont
  {Hiramatsu}}, \bibinfo {author} {\bibfnamefont {E.~I.}\ \bibnamefont
  {Sfakianakis}}, \ and\ \bibinfo {author} {\bibfnamefont {M.}~\bibnamefont
  {Yamaguchi}},\ }\href {\doibase 10.1007/JHEP03(2021)021} {\bibfield
  {journal} {\bibinfo  {journal} {JHEP}\ }\textbf {\bibinfo {volume} {03}},\
  \bibinfo {pages} {021} (\bibinfo {year} {2021})},\ \Eprint
  {http://arxiv.org/abs/2011.12201} {arXiv:2011.12201 [hep-ph]} \BibitemShut
  {NoStop}%
\bibitem [{\citenamefont {Cotner}\ \emph {et~al.}(2018)\citenamefont {Cotner},
  \citenamefont {Kusenko},\ and\ \citenamefont {Takhistov}}]{Cotner:2018vug}%
  \BibitemOpen
  \bibfield  {author} {\bibinfo {author} {\bibfnamefont {E.}~\bibnamefont
  {Cotner}}, \bibinfo {author} {\bibfnamefont {A.}~\bibnamefont {Kusenko}}, \
  and\ \bibinfo {author} {\bibfnamefont {V.}~\bibnamefont {Takhistov}},\ }\href
  {\doibase 10.1103/PhysRevD.98.083513} {\bibfield  {journal} {\bibinfo
  {journal} {Phys. Rev. D}\ }\textbf {\bibinfo {volume} {98}},\ \bibinfo
  {pages} {083513} (\bibinfo {year} {2018})},\ \Eprint
  {http://arxiv.org/abs/1801.03321} {arXiv:1801.03321 [astro-ph.CO]}
  \BibitemShut {NoStop}%
\bibitem [{\citenamefont {Cotner}\ \emph {et~al.}(2019)\citenamefont {Cotner},
  \citenamefont {Kusenko}, \citenamefont {Sasaki},\ and\ \citenamefont
  {Takhistov}}]{Cotner:2019ykd}%
  \BibitemOpen
  \bibfield  {author} {\bibinfo {author} {\bibfnamefont {E.}~\bibnamefont
  {Cotner}}, \bibinfo {author} {\bibfnamefont {A.}~\bibnamefont {Kusenko}},
  \bibinfo {author} {\bibfnamefont {M.}~\bibnamefont {Sasaki}}, \ and\ \bibinfo
  {author} {\bibfnamefont {V.}~\bibnamefont {Takhistov}},\ }\href {\doibase
  10.1088/1475-7516/2019/10/077} {\bibfield  {journal} {\bibinfo  {journal}
  {JCAP}\ }\textbf {\bibinfo {volume} {10}},\ \bibinfo {pages} {077} (\bibinfo
  {year} {2019})},\ \Eprint {http://arxiv.org/abs/1907.10613} {arXiv:1907.10613
  [astro-ph.CO]} \BibitemShut {NoStop}%
\bibitem [{sup()}]{supmat}%
  \BibitemOpen
  \href@noop {} {}\bibinfo {howpublished} {See Supplemental Material [url] for
  parameters of inflationary models, instability analysis and expressions for
  induced gravitational waves, which includes
  Refs.~\cite{Planck:2018jri,Amin:2011hj,
  Lozanov:2016hid,Lozanov:2017hjm,Lozanov:2014zfa,Inomata:2016rbd,Kohri:2018awv}.}\BibitemShut
  {Stop}%
\bibitem [{\citenamefont {Dvali}\ \emph {et~al.}(1994)\citenamefont {Dvali},
  \citenamefont {Shafi},\ and\ \citenamefont {Schaefer}}]{Dvali:1994ms}%
  \BibitemOpen
  \bibfield  {author} {\bibinfo {author} {\bibfnamefont {G.~R.}\ \bibnamefont
  {Dvali}}, \bibinfo {author} {\bibfnamefont {Q.}~\bibnamefont {Shafi}}, \ and\
  \bibinfo {author} {\bibfnamefont {R.~K.}\ \bibnamefont {Schaefer}},\ }\href
  {\doibase 10.1103/PhysRevLett.73.1886} {\bibfield  {journal} {\bibinfo
  {journal} {Phys. Rev. Lett.}\ }\textbf {\bibinfo {volume} {73}},\ \bibinfo
  {pages} {1886} (\bibinfo {year} {1994})},\ \Eprint
  {http://arxiv.org/abs/hep-ph/9406319} {arXiv:hep-ph/9406319} \BibitemShut
  {NoStop}%
\bibitem [{\citenamefont {Lee}\ and\ \citenamefont {Pang}(1992)}]{Lee:1991ax}%
  \BibitemOpen
  \bibfield  {author} {\bibinfo {author} {\bibfnamefont {T.~D.}\ \bibnamefont
  {Lee}}\ and\ \bibinfo {author} {\bibfnamefont {Y.}~\bibnamefont {Pang}},\
  }\href {\doibase 10.1016/0370-1573(92)90064-7} {\bibfield  {journal}
  {\bibinfo  {journal} {Phys. Rept.}\ }\textbf {\bibinfo {volume} {221}},\
  \bibinfo {pages} {251} (\bibinfo {year} {1992})}\BibitemShut {NoStop}%
\bibitem [{\citenamefont {Amin}(2013)}]{Amin:2013ika}%
  \BibitemOpen
  \bibfield  {author} {\bibinfo {author} {\bibfnamefont {M.~A.}\ \bibnamefont
  {Amin}},\ }\href {\doibase 10.1103/PhysRevD.87.123505} {\bibfield  {journal}
  {\bibinfo  {journal} {Phys. Rev. D}\ }\textbf {\bibinfo {volume} {87}},\
  \bibinfo {pages} {123505} (\bibinfo {year} {2013})},\ \Eprint
  {http://arxiv.org/abs/1303.1102} {arXiv:1303.1102 [astro-ph.CO]} \BibitemShut
  {NoStop}%
\bibitem [{\citenamefont {Mukaida}\ \emph {et~al.}(2017)\citenamefont
  {Mukaida}, \citenamefont {Takimoto},\ and\ \citenamefont
  {Yamada}}]{Mukaida:2016hwd}%
  \BibitemOpen
  \bibfield  {author} {\bibinfo {author} {\bibfnamefont {K.}~\bibnamefont
  {Mukaida}}, \bibinfo {author} {\bibfnamefont {M.}~\bibnamefont {Takimoto}}, \
  and\ \bibinfo {author} {\bibfnamefont {M.}~\bibnamefont {Yamada}},\ }\href
  {\doibase 10.1007/JHEP03(2017)122} {\bibfield  {journal} {\bibinfo  {journal}
  {JHEP}\ }\textbf {\bibinfo {volume} {03}},\ \bibinfo {pages} {122} (\bibinfo
  {year} {2017})},\ \Eprint {http://arxiv.org/abs/1612.07750} {arXiv:1612.07750
  [hep-ph]} \BibitemShut {NoStop}%
\bibitem [{\citenamefont {Ibe}\ \emph {et~al.}(2019)\citenamefont {Ibe},
  \citenamefont {Kawasaki}, \citenamefont {Nakano},\ and\ \citenamefont
  {Sonomoto}}]{Ibe:2019vyo}%
  \BibitemOpen
  \bibfield  {author} {\bibinfo {author} {\bibfnamefont {M.}~\bibnamefont
  {Ibe}}, \bibinfo {author} {\bibfnamefont {M.}~\bibnamefont {Kawasaki}},
  \bibinfo {author} {\bibfnamefont {W.}~\bibnamefont {Nakano}}, \ and\ \bibinfo
  {author} {\bibfnamefont {E.}~\bibnamefont {Sonomoto}},\ }\href {\doibase
  10.1007/JHEP04(2019)030} {\bibfield  {journal} {\bibinfo  {journal} {JHEP}\
  }\textbf {\bibinfo {volume} {04}},\ \bibinfo {pages} {030} (\bibinfo {year}
  {2019})},\ \Eprint {http://arxiv.org/abs/1901.06130} {arXiv:1901.06130
  [hep-ph]} \BibitemShut {NoStop}%
\bibitem [{\citenamefont {Zhang}\ \emph {et~al.}(2020)\citenamefont {Zhang},
  \citenamefont {Amin}, \citenamefont {Copeland}, \citenamefont {Saffin},\ and\
  \citenamefont {Lozanov}}]{Zhang:2020bec}%
  \BibitemOpen
  \bibfield  {author} {\bibinfo {author} {\bibfnamefont {H.-Y.}\ \bibnamefont
  {Zhang}}, \bibinfo {author} {\bibfnamefont {M.~A.}\ \bibnamefont {Amin}},
  \bibinfo {author} {\bibfnamefont {E.~J.}\ \bibnamefont {Copeland}}, \bibinfo
  {author} {\bibfnamefont {P.~M.}\ \bibnamefont {Saffin}}, \ and\ \bibinfo
  {author} {\bibfnamefont {K.~D.}\ \bibnamefont {Lozanov}},\ }\href {\doibase
  10.1088/1475-7516/2020/07/055} {\bibfield  {journal} {\bibinfo  {journal}
  {JCAP}\ }\textbf {\bibinfo {volume} {07}},\ \bibinfo {pages} {055} (\bibinfo
  {year} {2020})},\ \Eprint {http://arxiv.org/abs/2004.01202} {arXiv:2004.01202
  [hep-th]} \BibitemShut {NoStop}%
\bibitem [{\citenamefont {Oll\'e}\ \emph {et~al.}(2020)\citenamefont {Oll\'e},
  \citenamefont {Pujol\`as},\ and\ \citenamefont {Rompineve}}]{Olle:2019kbo}%
  \BibitemOpen
  \bibfield  {author} {\bibinfo {author} {\bibfnamefont {J.}~\bibnamefont
  {Oll\'e}}, \bibinfo {author} {\bibfnamefont {O.}~\bibnamefont {Pujol\`as}}, \
  and\ \bibinfo {author} {\bibfnamefont {F.}~\bibnamefont {Rompineve}},\ }\href
  {\doibase 10.1088/1475-7516/2020/02/006} {\bibfield  {journal} {\bibinfo
  {journal} {JCAP}\ }\textbf {\bibinfo {volume} {02}},\ \bibinfo {pages} {006}
  (\bibinfo {year} {2020})},\ \Eprint {http://arxiv.org/abs/1906.06352}
  {arXiv:1906.06352 [hep-ph]} \BibitemShut {NoStop}%
\bibitem [{\citenamefont {Antusch}\ \emph {et~al.}(2019)\citenamefont
  {Antusch}, \citenamefont {Cefal\`a},\ and\ \citenamefont
  {Torrent\'\i{}}}]{Antusch:2019qrr}%
  \BibitemOpen
  \bibfield  {author} {\bibinfo {author} {\bibfnamefont {S.}~\bibnamefont
  {Antusch}}, \bibinfo {author} {\bibfnamefont {F.}~\bibnamefont {Cefal\`a}}, \
  and\ \bibinfo {author} {\bibfnamefont {F.}~\bibnamefont {Torrent\'\i{}}},\
  }\href {\doibase 10.1088/1475-7516/2019/10/002} {\bibfield  {journal}
  {\bibinfo  {journal} {JCAP}\ }\textbf {\bibinfo {volume} {10}},\ \bibinfo
  {pages} {002} (\bibinfo {year} {2019})},\ \Eprint
  {http://arxiv.org/abs/1907.00611} {arXiv:1907.00611 [hep-ph]} \BibitemShut
  {NoStop}%
\bibitem [{\citenamefont {Alabidi}\ \emph {et~al.}(2013)\citenamefont
  {Alabidi}, \citenamefont {Kohri}, \citenamefont {Sasaki},\ and\ \citenamefont
  {Sendouda}}]{Alabidi:2013lya}%
  \BibitemOpen
  \bibfield  {author} {\bibinfo {author} {\bibfnamefont {L.}~\bibnamefont
  {Alabidi}}, \bibinfo {author} {\bibfnamefont {K.}~\bibnamefont {Kohri}},
  \bibinfo {author} {\bibfnamefont {M.}~\bibnamefont {Sasaki}}, \ and\ \bibinfo
  {author} {\bibfnamefont {Y.}~\bibnamefont {Sendouda}},\ }\href {\doibase
  10.1088/1475-7516/2013/05/033} {\bibfield  {journal} {\bibinfo  {journal}
  {JCAP}\ }\textbf {\bibinfo {volume} {05}},\ \bibinfo {pages} {033} (\bibinfo
  {year} {2013})},\ \Eprint {http://arxiv.org/abs/1303.4519} {arXiv:1303.4519
  [astro-ph.CO]} \BibitemShut {NoStop}%
\bibitem [{\citenamefont {Dom\`enech}\ \emph
  {et~al.}(2021{\natexlab{b}})\citenamefont {Dom\`enech}, \citenamefont {Lin},\
  and\ \citenamefont {Sasaki}}]{Domenech:2020ssp}%
  \BibitemOpen
  \bibfield  {author} {\bibinfo {author} {\bibfnamefont {G.}~\bibnamefont
  {Dom\`enech}}, \bibinfo {author} {\bibfnamefont {C.}~\bibnamefont {Lin}}, \
  and\ \bibinfo {author} {\bibfnamefont {M.}~\bibnamefont {Sasaki}},\ }\href
  {\doibase 10.1088/1475-7516/2021/11/E01} {\bibfield  {journal} {\bibinfo
  {journal} {JCAP}\ }\textbf {\bibinfo {volume} {04}},\ \bibinfo {pages} {062}
  (\bibinfo {year} {2021}{\natexlab{b}})},\ \bibinfo {note} {[Erratum: JCAP 11,
  E01 (2021)]},\ \Eprint {http://arxiv.org/abs/2012.08151} {arXiv:2012.08151
  [gr-qc]} \BibitemShut {NoStop}%
\bibitem [{\citenamefont {Papanikolaou}\ \emph {et~al.}(2021)\citenamefont
  {Papanikolaou}, \citenamefont {Vennin},\ and\ \citenamefont
  {Langlois}}]{Papanikolaou:2020qtd}%
  \BibitemOpen
  \bibfield  {author} {\bibinfo {author} {\bibfnamefont {T.}~\bibnamefont
  {Papanikolaou}}, \bibinfo {author} {\bibfnamefont {V.}~\bibnamefont
  {Vennin}}, \ and\ \bibinfo {author} {\bibfnamefont {D.}~\bibnamefont
  {Langlois}},\ }\href {\doibase 10.1088/1475-7516/2021/03/053} {\bibfield
  {journal} {\bibinfo  {journal} {JCAP}\ }\textbf {\bibinfo {volume} {03}},\
  \bibinfo {pages} {053} (\bibinfo {year} {2021})},\ \Eprint
  {http://arxiv.org/abs/2010.11573} {arXiv:2010.11573 [astro-ph.CO]}
  \BibitemShut {NoStop}%
\bibitem [{\citenamefont {Passaglia}\ and\ \citenamefont
  {Sasaki}(2021)}]{Passaglia:2021jla}%
  \BibitemOpen
  \bibfield  {author} {\bibinfo {author} {\bibfnamefont {S.}~\bibnamefont
  {Passaglia}}\ and\ \bibinfo {author} {\bibfnamefont {M.}~\bibnamefont
  {Sasaki}},\ }\href@noop {} {\  (\bibinfo {year} {2021})},\ \Eprint
  {http://arxiv.org/abs/2109.12824} {arXiv:2109.12824 [astro-ph.CO]}
  \BibitemShut {NoStop}%
\bibitem [{\citenamefont {Dom\`enech}\ \emph {et~al.}(2022)\citenamefont
  {Dom\`enech}, \citenamefont {Passaglia},\ and\ \citenamefont
  {Renaux-Petel}}]{Domenech:2021and}%
  \BibitemOpen
  \bibfield  {author} {\bibinfo {author} {\bibfnamefont {G.}~\bibnamefont
  {Dom\`enech}}, \bibinfo {author} {\bibfnamefont {S.}~\bibnamefont
  {Passaglia}}, \ and\ \bibinfo {author} {\bibfnamefont {S.}~\bibnamefont
  {Renaux-Petel}},\ }\href {\doibase 10.1088/1475-7516/2022/03/023} {\bibfield
  {journal} {\bibinfo  {journal} {JCAP}\ }\textbf {\bibinfo {volume} {03}},\
  \bibinfo {pages} {023} (\bibinfo {year} {2022})},\ \Eprint
  {http://arxiv.org/abs/2112.10163} {arXiv:2112.10163 [astro-ph.CO]}
  \BibitemShut {NoStop}%
\bibitem [{\citenamefont {Lozanov}\ \emph {et~al.}()\citenamefont {Lozanov},
  \citenamefont {Sasaki},\ and\ \citenamefont {Takhistov}}]{Lozanov:2022fut}%
  \BibitemOpen
  \bibfield  {author} {\bibinfo {author} {\bibfnamefont {K.}~\bibnamefont
  {Lozanov}}, \bibinfo {author} {\bibfnamefont {M.}~\bibnamefont {Sasaki}}, \
  and\ \bibinfo {author} {\bibfnamefont {V.}~\bibnamefont {Takhistov}},\
  }\href@noop {} {\bibinfo  {journal} {\textit{in preparation}}\ }\BibitemShut
  {NoStop}%
\bibitem [{\citenamefont {Inomata}\ \emph {et~al.}(2020)\citenamefont
  {Inomata}, \citenamefont {Kawasaki}, \citenamefont {Mukaida}, \citenamefont
  {Terada},\ and\ \citenamefont {Yanagida}}]{Inomata:2020lmk}%
  \BibitemOpen
\bibfield  {journal} {  }\bibfield  {author} {\bibinfo {author} {\bibfnamefont
  {K.}~\bibnamefont {Inomata}}, \bibinfo {author} {\bibfnamefont
  {M.}~\bibnamefont {Kawasaki}}, \bibinfo {author} {\bibfnamefont
  {K.}~\bibnamefont {Mukaida}}, \bibinfo {author} {\bibfnamefont
  {T.}~\bibnamefont {Terada}}, \ and\ \bibinfo {author} {\bibfnamefont {T.~T.}\
  \bibnamefont {Yanagida}},\ }\href {\doibase 10.1103/PhysRevD.101.123533}
  {\bibfield  {journal} {\bibinfo  {journal} {Phys. Rev. D}\ }\textbf {\bibinfo
  {volume} {101}},\ \bibinfo {pages} {123533} (\bibinfo {year} {2020})},\
  \Eprint {http://arxiv.org/abs/2003.10455} {arXiv:2003.10455 [astro-ph.CO]}
  \BibitemShut {NoStop}%
\bibitem [{\citenamefont {Hertzberg}(2010)}]{Hertzberg:2010yz}%
  \BibitemOpen
  \bibfield  {author} {\bibinfo {author} {\bibfnamefont {M.~P.}\ \bibnamefont
  {Hertzberg}},\ }\href {\doibase 10.1103/PhysRevD.82.045022} {\bibfield
  {journal} {\bibinfo  {journal} {Phys. Rev. D}\ }\textbf {\bibinfo {volume}
  {82}},\ \bibinfo {pages} {045022} (\bibinfo {year} {2010})},\ \Eprint
  {http://arxiv.org/abs/1003.3459} {arXiv:1003.3459 [hep-th]} \BibitemShut
  {NoStop}%
\bibitem [{\citenamefont {Kohri}\ and\ \citenamefont
  {Terada}(2018)}]{Kohri:2018awv}%
  \BibitemOpen
  \bibfield  {author} {\bibinfo {author} {\bibfnamefont {K.}~\bibnamefont
  {Kohri}}\ and\ \bibinfo {author} {\bibfnamefont {T.}~\bibnamefont {Terada}},\
  }\href {\doibase 10.1103/PhysRevD.97.123532} {\bibfield  {journal} {\bibinfo
  {journal} {Phys. Rev. D}\ }\textbf {\bibinfo {volume} {97}},\ \bibinfo
  {pages} {123532} (\bibinfo {year} {2018})},\ \Eprint
  {http://arxiv.org/abs/1804.08577} {arXiv:1804.08577 [gr-qc]} \BibitemShut
  {NoStop}%
\bibitem [{\citenamefont {Inomata}\ \emph {et~al.}(2017)\citenamefont
  {Inomata}, \citenamefont {Kawasaki}, \citenamefont {Mukaida}, \citenamefont
  {Tada},\ and\ \citenamefont {Yanagida}}]{Inomata:2016rbd}%
  \BibitemOpen
  \bibfield  {author} {\bibinfo {author} {\bibfnamefont {K.}~\bibnamefont
  {Inomata}}, \bibinfo {author} {\bibfnamefont {M.}~\bibnamefont {Kawasaki}},
  \bibinfo {author} {\bibfnamefont {K.}~\bibnamefont {Mukaida}}, \bibinfo
  {author} {\bibfnamefont {Y.}~\bibnamefont {Tada}}, \ and\ \bibinfo {author}
  {\bibfnamefont {T.~T.}\ \bibnamefont {Yanagida}},\ }\href {\doibase
  10.1103/PhysRevD.95.123510} {\bibfield  {journal} {\bibinfo  {journal} {Phys.
  Rev. D}\ }\textbf {\bibinfo {volume} {95}},\ \bibinfo {pages} {123510}
  (\bibinfo {year} {2017})},\ \Eprint {http://arxiv.org/abs/1611.06130}
  {arXiv:1611.06130 [astro-ph.CO]} \BibitemShut {NoStop}%
\bibitem [{\citenamefont {Assadullahi}\ and\ \citenamefont
  {Wands}(2009)}]{Assadullahi:2009nf}%
  \BibitemOpen
  \bibfield  {author} {\bibinfo {author} {\bibfnamefont {H.}~\bibnamefont
  {Assadullahi}}\ and\ \bibinfo {author} {\bibfnamefont {D.}~\bibnamefont
  {Wands}},\ }\href {\doibase 10.1103/PhysRevD.79.083511} {\bibfield  {journal}
  {\bibinfo  {journal} {Phys. Rev. D}\ }\textbf {\bibinfo {volume} {79}},\
  \bibinfo {pages} {083511} (\bibinfo {year} {2009})},\ \Eprint
  {http://arxiv.org/abs/0901.0989} {arXiv:0901.0989 [astro-ph.CO]} \BibitemShut
  {NoStop}%
\bibitem [{\citenamefont {Aghanim}\ \emph {et~al.}(2020)\citenamefont {Aghanim}
  \emph {et~al.}}]{Planck:2018vyg}%
  \BibitemOpen
  \bibfield  {author} {\bibinfo {author} {\bibfnamefont {N.}~\bibnamefont
  {Aghanim}} \emph {et~al.} (\bibinfo {collaboration} {Planck}),\ }\href
  {\doibase 10.1051/0004-6361/201833910} {\bibfield  {journal} {\bibinfo
  {journal} {Astron. Astrophys.}\ }\textbf {\bibinfo {volume} {641}},\ \bibinfo
  {pages} {A6} (\bibinfo {year} {2020})},\ \bibinfo {note} {[Erratum:
  Astron.Astrophys. 652, C4 (2021)]},\ \Eprint
  {http://arxiv.org/abs/1807.06209} {arXiv:1807.06209 [astro-ph.CO]}
  \BibitemShut {NoStop}%
\bibitem [{\citenamefont {Abazajian}\ \emph {et~al.}(2019)\citenamefont
  {Abazajian} \emph {et~al.}}]{Abazajian:2019eic}%
  \BibitemOpen
  \bibfield  {author} {\bibinfo {author} {\bibfnamefont {K.}~\bibnamefont
  {Abazajian}} \emph {et~al.},\ }\href@noop {} {\  (\bibinfo {year} {2019})},\
  \Eprint {http://arxiv.org/abs/1907.04473} {arXiv:1907.04473 [astro-ph.IM]}
  \BibitemShut {NoStop}%
\bibitem [{\citenamefont {Schmitz}(2021)}]{Schmitz:2020syl}%
  \BibitemOpen
  \bibfield  {author} {\bibinfo {author} {\bibfnamefont {K.}~\bibnamefont
  {Schmitz}},\ }\href {\doibase 10.1007/JHEP01(2021)097} {\bibfield  {journal}
  {\bibinfo  {journal} {JHEP}\ }\textbf {\bibinfo {volume} {01}},\ \bibinfo
  {pages} {097} (\bibinfo {year} {2021})},\ \Eprint
  {http://arxiv.org/abs/2002.04615} {arXiv:2002.04615 [hep-ph]} \BibitemShut
  {NoStop}%
\bibitem [{\citenamefont {Abbott}\ \emph {et~al.}(2016)\citenamefont {Abbott}
  \emph {et~al.}}]{LIGOScientific:2016fpe}%
  \BibitemOpen
  \bibfield  {author} {\bibinfo {author} {\bibfnamefont {B.~P.}\ \bibnamefont
  {Abbott}} \emph {et~al.} (\bibinfo {collaboration} {LIGO Scientific,
  Virgo}),\ }\href {\doibase 10.1103/PhysRevLett.116.131102} {\bibfield
  {journal} {\bibinfo  {journal} {Phys. Rev. Lett.}\ }\textbf {\bibinfo
  {volume} {116}},\ \bibinfo {pages} {131102} (\bibinfo {year} {2016})},\
  \Eprint {http://arxiv.org/abs/1602.03847} {arXiv:1602.03847 [gr-qc]}
  \BibitemShut {NoStop}%
\bibitem [{\citenamefont {Birrell}\ and\ \citenamefont
  {Davies}(1984)}]{Birrell:1982ix}%
  \BibitemOpen
  \bibfield  {author} {\bibinfo {author} {\bibfnamefont {N.~D.}\ \bibnamefont
  {Birrell}}\ and\ \bibinfo {author} {\bibfnamefont {P.~C.~W.}\ \bibnamefont
  {Davies}},\ }\href {\doibase 10.1017/CBO9780511622632} {\emph {\bibinfo
  {title} {{Quantum Fields in Curved Space}}}},\ Cambridge Monographs on
  Mathematical Physics\ (\bibinfo  {publisher} {Cambridge Univ. Press},\
  \bibinfo {address} {Cambridge, UK},\ \bibinfo {year} {1984})\BibitemShut
  {NoStop}%
\bibitem [{\citenamefont {Adshead}\ \emph
  {et~al.}(2020{\natexlab{a}})\citenamefont {Adshead}, \citenamefont {Giblin},
  \citenamefont {Pieroni},\ and\ \citenamefont {Weiner}}]{Adshead:2019lbr}%
  \BibitemOpen
  \bibfield  {author} {\bibinfo {author} {\bibfnamefont {P.}~\bibnamefont
  {Adshead}}, \bibinfo {author} {\bibfnamefont {J.~T.}\ \bibnamefont {Giblin}},
  \bibinfo {author} {\bibfnamefont {M.}~\bibnamefont {Pieroni}}, \ and\
  \bibinfo {author} {\bibfnamefont {Z.~J.}\ \bibnamefont {Weiner}},\ }\href
  {\doibase 10.1103/PhysRevD.101.083534} {\bibfield  {journal} {\bibinfo
  {journal} {Phys. Rev. D}\ }\textbf {\bibinfo {volume} {101}},\ \bibinfo
  {pages} {083534} (\bibinfo {year} {2020}{\natexlab{a}})},\ \Eprint
  {http://arxiv.org/abs/1909.12842} {arXiv:1909.12842 [astro-ph.CO]}
  \BibitemShut {NoStop}%
\bibitem [{\citenamefont {Adshead}\ \emph
  {et~al.}(2020{\natexlab{b}})\citenamefont {Adshead}, \citenamefont {Giblin},
  \citenamefont {Pieroni},\ and\ \citenamefont {Weiner}}]{Adshead:2019igv}%
  \BibitemOpen
  \bibfield  {author} {\bibinfo {author} {\bibfnamefont {P.}~\bibnamefont
  {Adshead}}, \bibinfo {author} {\bibfnamefont {J.~T.}\ \bibnamefont {Giblin}},
  \bibinfo {author} {\bibfnamefont {M.}~\bibnamefont {Pieroni}}, \ and\
  \bibinfo {author} {\bibfnamefont {Z.~J.}\ \bibnamefont {Weiner}},\ }\href
  {\doibase 10.1103/PhysRevLett.124.171301} {\bibfield  {journal} {\bibinfo
  {journal} {Phys. Rev. Lett.}\ }\textbf {\bibinfo {volume} {124}},\ \bibinfo
  {pages} {171301} (\bibinfo {year} {2020}{\natexlab{b}})},\ \Eprint
  {http://arxiv.org/abs/1909.12843} {arXiv:1909.12843 [astro-ph.CO]}
  \BibitemShut {NoStop}%
\bibitem [{\citenamefont {Aggarwal}\ \emph {et~al.}(2021)\citenamefont
  {Aggarwal} \emph {et~al.}}]{Aggarwal:2020olq}%
  \BibitemOpen
  \bibfield  {author} {\bibinfo {author} {\bibfnamefont {N.}~\bibnamefont
  {Aggarwal}} \emph {et~al.},\ }\href {\doibase 10.1007/s41114-021-00032-5}
  {\bibfield  {journal} {\bibinfo  {journal} {Living Rev. Rel.}\ }\textbf
  {\bibinfo {volume} {24}},\ \bibinfo {pages} {4} (\bibinfo {year} {2021})},\
  \Eprint {http://arxiv.org/abs/2011.12414} {arXiv:2011.12414 [gr-qc]}
  \BibitemShut {NoStop}%
\bibitem [{\citenamefont {Caprini}\ and\ \citenamefont
  {Figueroa}(2018)}]{Caprini:2018mtu}%
  \BibitemOpen
  \bibfield  {author} {\bibinfo {author} {\bibfnamefont {C.}~\bibnamefont
  {Caprini}}\ and\ \bibinfo {author} {\bibfnamefont {D.~G.}\ \bibnamefont
  {Figueroa}},\ }\href {\doibase 10.1088/1361-6382/aac608} {\bibfield
  {journal} {\bibinfo  {journal} {Class. Quant. Grav.}\ }\textbf {\bibinfo
  {volume} {35}},\ \bibinfo {pages} {163001} (\bibinfo {year} {2018})},\
  \Eprint {http://arxiv.org/abs/1801.04268} {arXiv:1801.04268 [astro-ph.CO]}
  \BibitemShut {NoStop}%
\bibitem [{\citenamefont {Lozanov}\ and\ \citenamefont
  {Amin}(2017)}]{Lozanov:2016hid}%
  \BibitemOpen
  \bibfield  {author} {\bibinfo {author} {\bibfnamefont {K.~D.}\ \bibnamefont
  {Lozanov}}\ and\ \bibinfo {author} {\bibfnamefont {M.~A.}\ \bibnamefont
  {Amin}},\ }\href {\doibase 10.1103/PhysRevLett.119.061301} {\bibfield
  {journal} {\bibinfo  {journal} {Phys. Rev. Lett.}\ }\textbf {\bibinfo
  {volume} {119}},\ \bibinfo {pages} {061301} (\bibinfo {year} {2017})},\
  \Eprint {http://arxiv.org/abs/1608.01213} {arXiv:1608.01213 [astro-ph.CO]}
  \BibitemShut {NoStop}%
\bibitem [{\citenamefont {Lozanov}\ and\ \citenamefont
  {Amin}(2018)}]{Lozanov:2017hjm}%
  \BibitemOpen
  \bibfield  {author} {\bibinfo {author} {\bibfnamefont {K.~D.}\ \bibnamefont
  {Lozanov}}\ and\ \bibinfo {author} {\bibfnamefont {M.~A.}\ \bibnamefont
  {Amin}},\ }\href {\doibase 10.1103/PhysRevD.97.023533} {\bibfield  {journal}
  {\bibinfo  {journal} {Phys. Rev. D}\ }\textbf {\bibinfo {volume} {97}},\
  \bibinfo {pages} {023533} (\bibinfo {year} {2018})},\ \Eprint
  {http://arxiv.org/abs/1710.06851} {arXiv:1710.06851 [astro-ph.CO]}
  \BibitemShut {NoStop}%
\bibitem [{\citenamefont {Lozanov}\ and\ \citenamefont
  {Amin}(2014)}]{Lozanov:2014zfa}%
  \BibitemOpen
  \bibfield  {author} {\bibinfo {author} {\bibfnamefont {K.~D.}\ \bibnamefont
  {Lozanov}}\ and\ \bibinfo {author} {\bibfnamefont {M.~A.}\ \bibnamefont
  {Amin}},\ }\href {\doibase 10.1103/PhysRevD.90.083528} {\bibfield  {journal}
  {\bibinfo  {journal} {Phys. Rev. D}\ }\textbf {\bibinfo {volume} {90}},\
  \bibinfo {pages} {083528} (\bibinfo {year} {2014})},\ \Eprint
  {http://arxiv.org/abs/1408.1811} {arXiv:1408.1811 [hep-ph]} \BibitemShut
  {NoStop}%
\end{thebibliography}%

\clearpage
\onecolumngrid
\begin{center}
   \textbf{\large SUPPLEMENTAL MATERIAL \\[.1cm] ``Enhanced Gravitational Waves from Inflaton Oscillons''}\\[.2cm]
  \vspace{0.05in}
  {Kaloian D. Lozanov, Volodymyr Takhistov}
\end{center}

\twocolumngrid
\setcounter{equation}{0}
\setcounter{figure}{0}
\setcounter{table}{0}
\setcounter{section}{0}
\setcounter{page}{1}
\makeatletter
\renewcommand{\theequation}{S\arabic{equation}}
\renewcommand{\thefigure}{S\arabic{figure}}
\renewcommand{\thetable}{S\arabic{table}}

\onecolumngrid

In this Supplemental Material we provide detailed expressions for the inflationary model parameters of the scenarios
considered in the main text. Further, we perform instability analysis for the discussed potentials and give an overview
of the analytic expressions for induced gravitational waves associated with sudden matter-to-radiation epoch transition
for the generic power spectrum we consider in the main text.

\section*{Inflationary Model Parameters}
\label{sec:infmodels}

Observables related to CMB are directly linked to the inflaton potential~\cite{Planck:2018jri}
\begin{equation*} 
\left\{  
        \begin{array}{l}  
            A_{\rm s}=\dfrac{1}{12\pi^2}\dfrac{V_\star^3}{M_{\rm Pl}^6V_{\star}'^2}\approx2.1\times10^{-9}\,,\\
            n_{\rm s}=1-3M_{\rm Pl}^2\dfrac{V_\star'^2}{V_\star^2}+2M_{\rm Pl}^2\dfrac{V_\star''}{V_\star}=0.9668\pm0.0037\,,\\
            r=8M_{\rm Pl}^2\dfrac{V_\star'^2}{V_\star^2}<0.063\,,\quad95\%\rm{\,\,CL}\,,\\
            N_\star=\left|\int_{\phi_\star}^{\phi_{\rm end}}{\dfrac{V_\star}{V_\star'}}\dfrac{d\phi}{M_{\rm Pl}^2}\right|\,,\\
        \end{array}
        \right.  
    \end{equation*}
where $V_\star\equiv V(\phi_\star)$, etc and $\phi_\star$ is the value of the inflaton field when the pivot scale crossed outside the Hubble sphere, $k=k_\star=a_\star H_\star$ with scale factor $a$ and Hubble parameter $H = a \mathcal{H}$, and $\phi_{\rm end}$ is the value of the inflaton at the end of inflation, when $\ddot{a}=0$.

Motivated potentials favored by CMB observations~\cite{Planck:2018jri} described in the main text include monodromy, log-normal
and pure natural (plateau) models. Their behavior relative to quadratic potential is shown on Fig.~1 of the main text.
Below we derive expressions for inflationary observables of these potentials, assuming slow-roll and $|\phi_\star|\gg|\phi_{\rm end}|$ for simplicity. For monodromy potential these are given by
these are given by 
\begin{equation}
     V(\phi) = m^2M^2 \left[\sqrt{1+\dfrac{\phi^2}{M^2}} - 1\right] ~~~~~\Longrightarrow~~~~~
     \left\{ 
        \begin{array}{l} 
            A_{\rm s} \approx \dfrac{1}{12\pi^2}\dfrac{m^2M}{M_{\rm Pl}^3}(2N_\star)^{3/2}\,,\\
            n_{\rm s}\approx1-\dfrac{3}{2N_\star}\,,\\
            r\approx\dfrac{4}{N_\star}\,,\\
N_\star\approx\dfrac{\phi_\star^2}{2M_{\rm Pl}^2},\\
        \end{array}
        \right. 
\end{equation} 

For log-normal potential these are given by
\begin{equation}
      V(\phi) = \frac{m^2M^2}{2} \ln\left(1+\frac{\phi^2}{M^2}\right) ~~~~~\Longrightarrow~~~~~
\left\{
        \begin{array}{l} 
            A_{\rm s} \approx \dfrac{1}{24\pi^2}\dfrac{m^2M^2}{M_{\rm Pl}^4}N_\star\ln^{2}\dfrac{\phi_\star^2}{M^2}\,,\\
            n_{\rm s}\approx1-\dfrac{1}{N_\star}\,,\\
            r\approx\dfrac{8}{N_\star}\dfrac{1}{\ln(\phi_\star^2/M^2)}\,,\\
N_\star\approx\dfrac{\phi_\star^2}{4M_{\rm Pl}^2}\ln\dfrac{\phi_\star^2}{M^2}\,,\\
        \end{array}
        \right. 
\end{equation} 
 
For pure natural (plateau) potentials of these are given by
 \begin{equation}
      V(\phi) = m^2M^2 \left[1-\frac{1}{\sqrt{1+\phi^2/M^2}}\right] ~~~~~\Longrightarrow~~~~~
\left\{ 
        \begin{array}{l} 
            A_{\rm s} \approx \dfrac{1}{12\pi^2}\dfrac{m^2}{M_{\rm Pl}^2}\left(3N_\star\dfrac{M}{M_{\rm Pl}}\right)^{4/3}\,,\\
            n_{\rm s}\approx1-\dfrac{4}{3N_\star}\,,\\
            r\approx\dfrac{8}{(3N_\star)^{4/3}}\left(\dfrac{M}{M_{\rm Pl}}\right)^{2/3}\,,\\
N_\star\approx\dfrac{\phi_\star^3}{3M_{\rm Pl}^2M}\,.\\
        \end{array}
        \right.
\end{equation} 
Here all the relevant variables follow the definitions of the main text.

\section*{FLOQUET STABILITY ANALYSIS}

Instabilities and formation of oscillons can be examined for models in question through application of the Floquet
analysis. For Floquet exponents, efficient resonance occurs when $|{\Re}(\mu_k)|/H \gtrsim \mathcal{O}(10)$.
This is also a prerequisite for
oscillon formation. Momentum modes $\delta \phi_k$ have physical wavenumber $k/a$, where $a$ is the expansion scale factor. They scan the Floquet bands as the scale factor changes with time. 

\begin{figure*}[h] 
   \centering
       \includegraphics[width=0.25\linewidth]{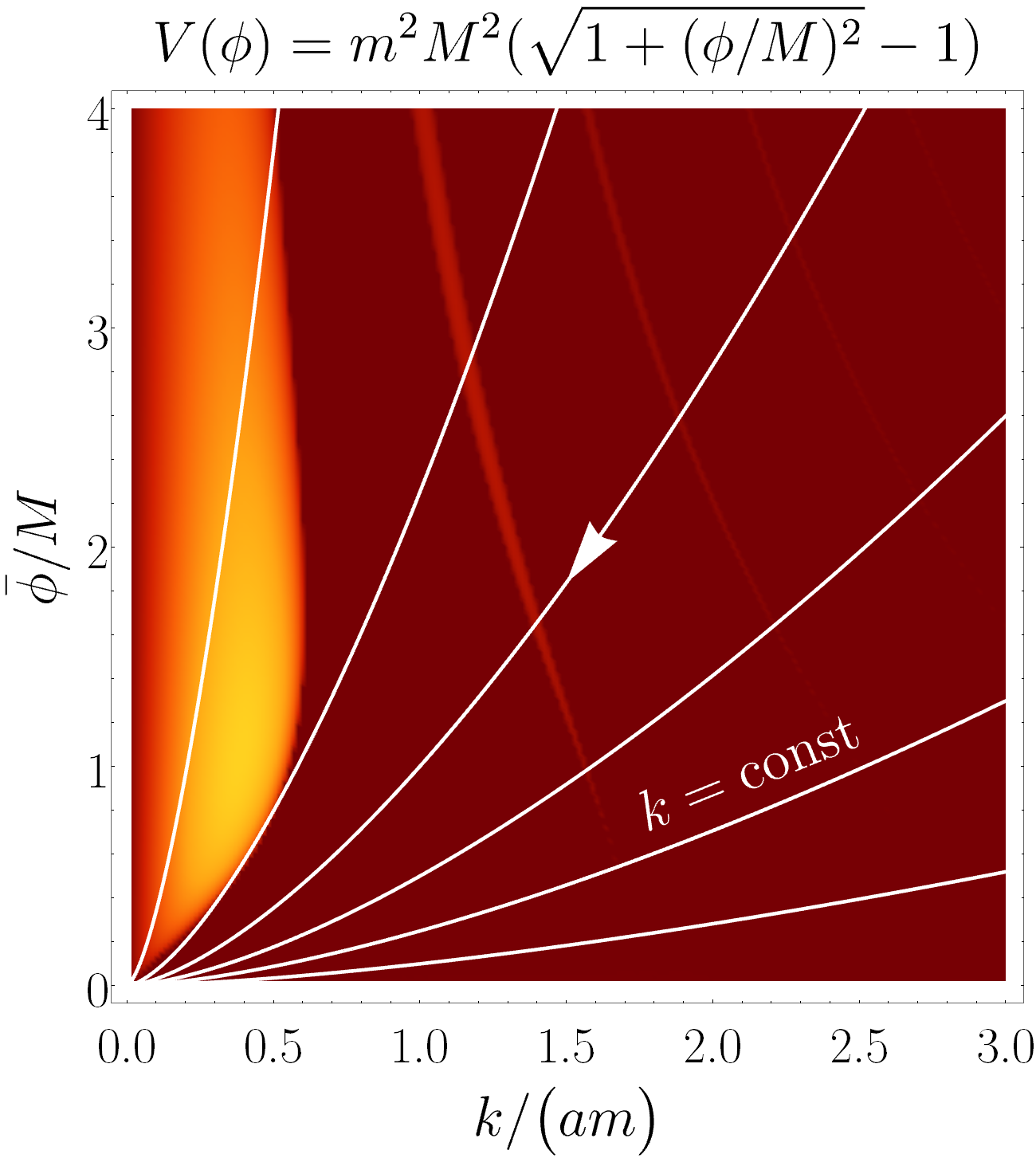}
        \raisebox{0.13\height}{\includegraphics[height=1.32in]{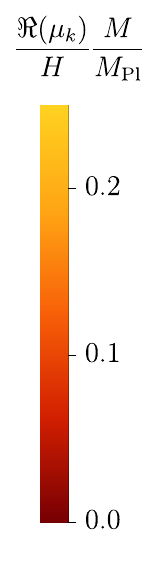}}
   \includegraphics[width=0.25\linewidth]{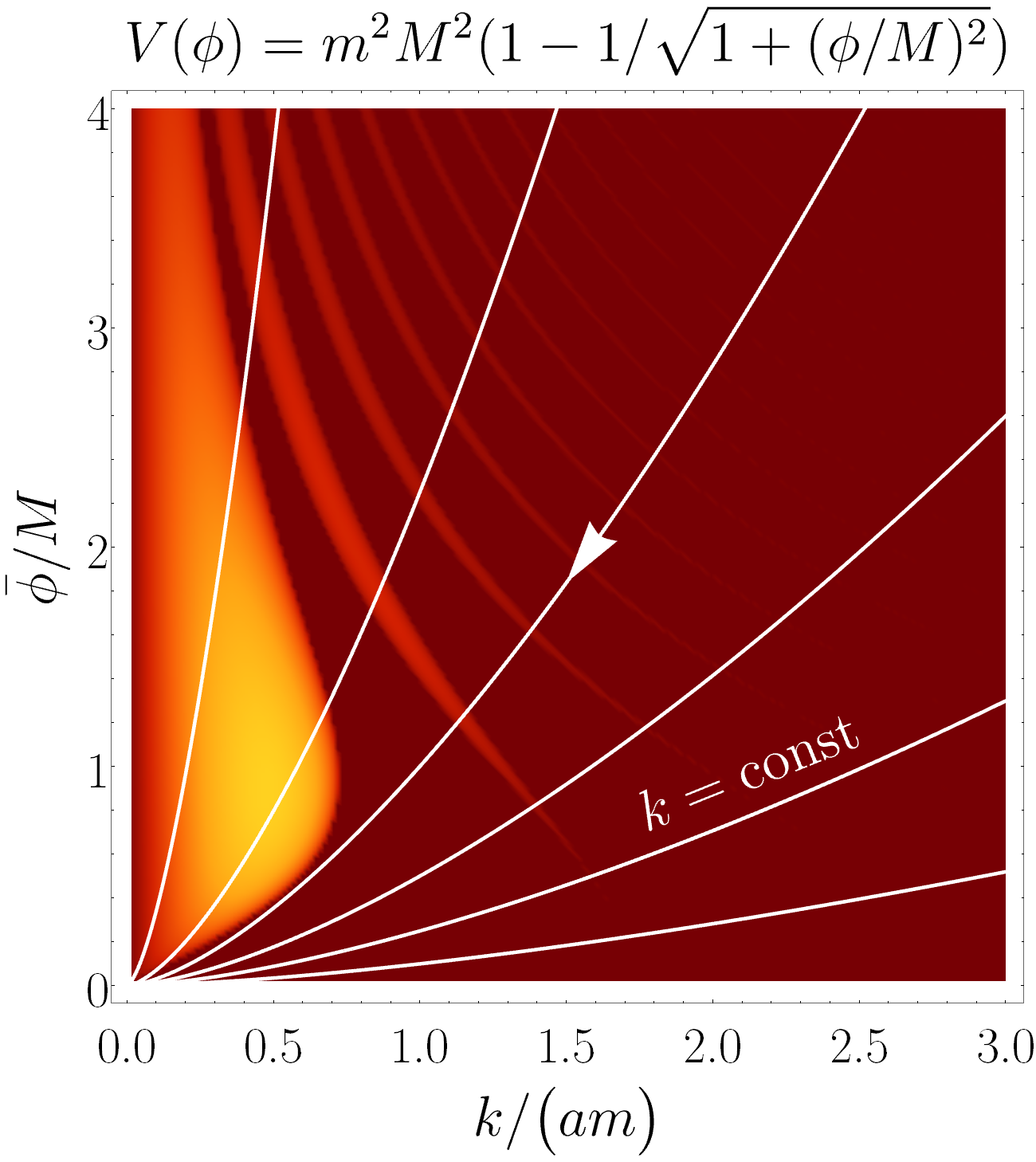} 
     \raisebox{0.13\height}{\includegraphics[height=1.32in]{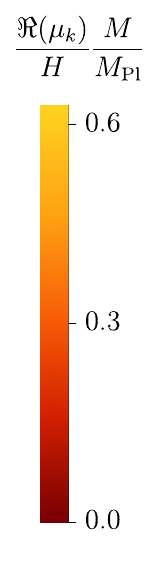}}
    \includegraphics[width=0.25\linewidth]{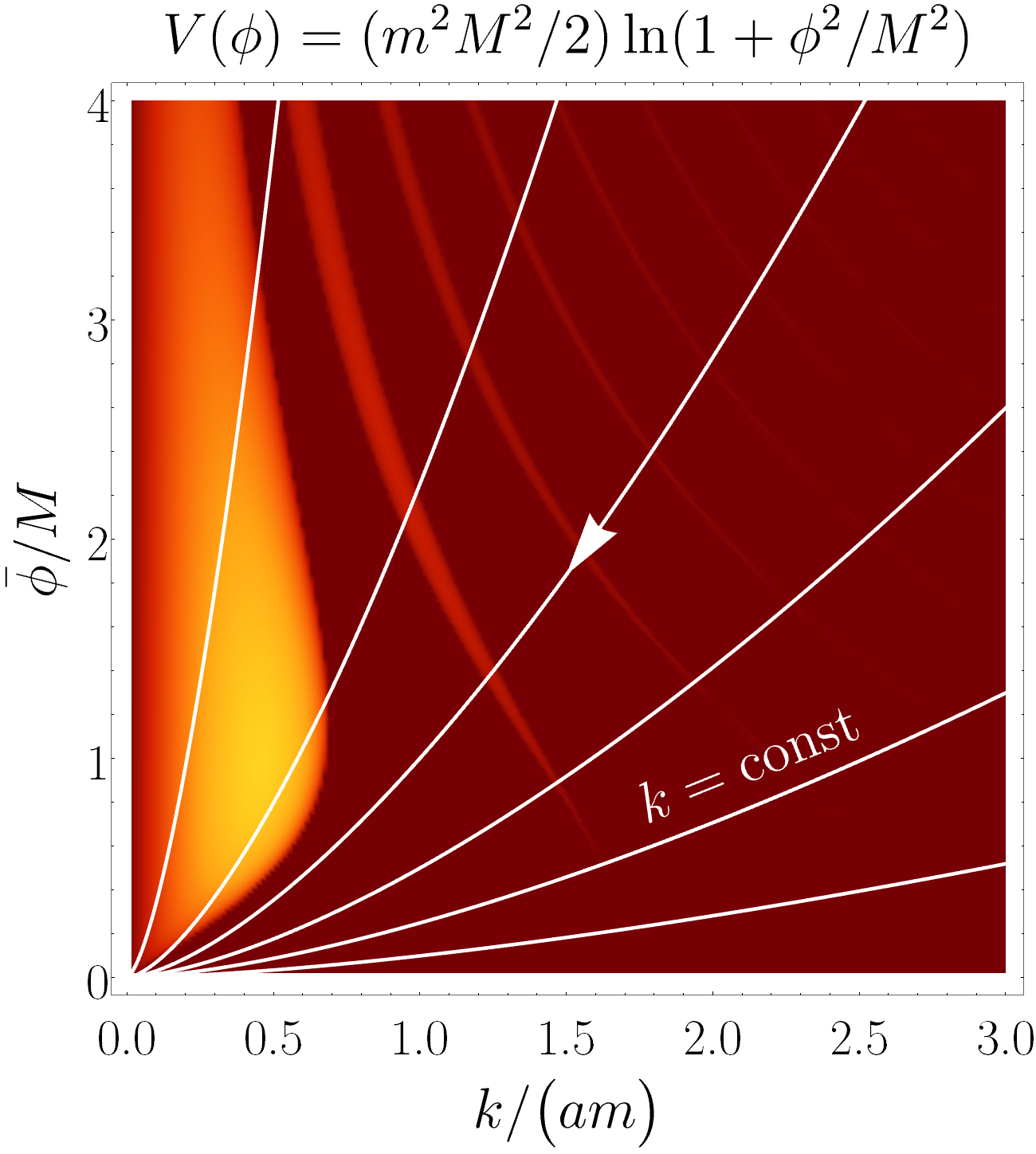} 
     \raisebox{0.13\height}{\includegraphics[height=1.32in]{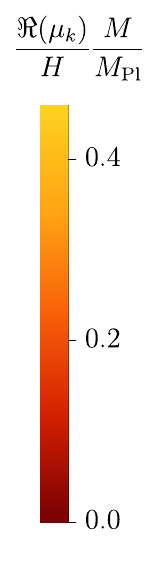}}
     \caption{Floquet charts for instability bands of scalar field fluctuations assuming monodromy model [\textbf{Left}], pure natural  (plateau)-inflation [\textbf{Middle}] and log-normal model [\textbf{Right}].  The vertical axis denotes the amplitude of oscillation of the homogeneous
mode $\bar{\phi}$. Lighter colors correspond to unstable regions. We set $H^2=V(\bar{\phi})/(3M_{\rm Pl}^2)$. The white lines depict qualitatively the evolution of modes of fixed co-moving wavenumber $k$ as time goes by and the Universe expands, causing $a$ to grow and $\bar{\phi}$ to decay. Efficient resonance takes place in FRW, provided $\Re(\mu_k)/H\gtrsim\mathcal{O}(10)$, implying $M\lesssim10^{-2}M_{\rm Pl}$.} 
   \label{fig:FloqH}   
\end{figure*}

We explicitly analyze the instability bands of potentials described in the main text and numerically compute their Floquet charts following Refs.~\cite{Amin:2011hj, Lozanov:2016hid,Lozanov:2017hjm,Lozanov:2014zfa}.
Results for monodoromy potential, pure natural (plateu) potential and log-normal
potential are depicted in Fig.~\ref{fig:FloqH}. Instability bands leading to oscillon formation are readily seen.

\section*{Analytic Expressions for Induced Gravitational Waves}
\label{sec:inducedgws}

Following Ref.~\cite{Inomata:2016rbd,Kohri:2018awv}, we provide an overview of the analytic expressions describing induced GWs associated with sudden matter-domination transition to radiation-domination and the generic power spectrum of Eq.~\eqref{eq:powerspec}, valid for $n_s > -3/2$. 

The kernel function for the curvature (scalar) power spectrum described by
\begin{equation} \label{eq:powerspec}
    \mathcal{P}_{\zeta} = A_s \Theta(k_{\rm max} - k) \Big( 
    \dfrac{k}{k_{\star}}\Big)^{n_s-1}~,
\end{equation}
where $k_{\rm max}$ is the cutoff scale,
is given by 
\begin{equation}
I(u,v,k,\eta,\eta_{\rm R}) = \int^{x}_0 {\rm d}\bar{x} \frac{a(\bar \eta)}{a(\eta)} k G_k(\eta, \bar \eta) f(u,v,\bar x, x_{\rm R}),
\end{equation}
where $x_{R} = k \eta_{R}$.
$G_k(\eta,\bar \eta)$ is the Green function
\begin{equation}
G_k''(\eta, \bar \eta) + \left( k^2 - \frac{a''(\eta)}{a(\eta)} \right) G_k(\eta, \bar \eta) = \delta (\eta - \bar \eta),
\end{equation}
where the prime denotes conformal derivative.
The source function of the gravitational potential $\Phi$ is
\begin{equation}
f(u,v,\bar{x}, x_{\rm R})=~ \frac{3}{25(1+w)}  \left[ 2(5+3w) \Phi(u\bar{x})\Phi(v\bar{x})   +4 \mathcal{H}^{-1} \left(\Phi'(u\bar{x})\Phi(v\bar{x}) + \Phi(u\bar{x})\Phi'(v\bar{x})\right)  + 4 \mathcal{H}^{-2} \Phi'(u\bar{x})\Phi'(v\bar{x}) \right]
\end{equation}
where $w$ describes equation of state. Sudden transition will enhance the contributions dependent on $\Phi^{\prime}$.

The induced GWs contributions and the kernel can be separated into eMD and RD components. The total spectrum is predominantly produced in RD after reheating transition with 
\begin{equation}
\Omega_{\rm GW} \simeq \Omega_{\rm GW, RD} \simeq \Omega_{\rm GW, RD}^{\textrm{(LS)}} + \Omega_{\rm GW, RD}^{(\textrm{res})}~,
\end{equation}
which consists of contributions from large scales (LS) as well as amplified small-scale oscillation resonant peaks (res). These are given by
\begin{align}
\label{eq:omega_tilt_ls}
\Omega_{\text{GW,RD}}^{\text{(LS)}}\simeq &~ \frac{3 \left( 4 \text{Ci}\left(\frac{x_{\rm R}}{2}\right)^2 + \left( \pi - 2  \text{Si}\left(\frac{x_{\rm R}}{2}\right) \right)^2 \right) A_\text{s}^2  x_\text{max,R}^8 }{2^{17+2 n_\text{s}} \times 625 (3+2n_\text{s})} \left( \frac{2 x_{\text{max,R}}}{x_\text{R}}-1 \right)^{2 n_\text{s}} \left( \frac{x_\text{R}}{x_{*,\text{R}}} \right)^{2(n_s -1)} \nonumber \\
& \qquad \times \left( \widetilde\Omega_{\text{GW,RD}}^{\text{(LS,1)}} \Theta (x_{\text{max,R}} - x_\text{R}) + \widetilde\Omega_{\text{GW,RD}}^{\text{(LS,2)}} \Theta (x_\text{R} - x_{\text{max,R}} )  \right)  \Theta ( 2 x_{\text{max,R}} - x_\text{R})~,
\end{align}
where 
\begin{align}
\label{eq:omega_tilt_ls1}
\widetilde\Omega_\text{GW,RD}^\text{(LS,1)} =&~ \frac{1}{(2+n_\text{s}) (3 + n_\text{s}) (4 + n_\text{s}) (5 + 2 n_\text{s}) (7 + 2 n_\text{s})} \Big( 1536 - 6144 \widetilde{k} + (7168-1920 n_\text{s} -256 n_\text{s}^2 ) \widetilde{k}^2 \notag \\
&  +(5760 n_\text{s} + 768 n_\text{s}^2) \widetilde{k}^3 +(1328 n_\text{s} + 3056 n_\text{s}^2 + 832 n_\text{s}^3 +64 n_\text{s}^4) \widetilde{k}^4  \nonumber \\
& - (7168+12256 n_\text{s} + 7392 n_\text{s}^2 +1664 n_\text{s}^3 + 128 n_\text{s}^4 ) \widetilde{k}^5 +(7392 + 10992 n_\text{s} + 5784 n_\text{s}^2 + 1248 n_\text{s}^3 + 96 n_\text{s}^4) \widetilde{k}^6   \nonumber   \\
&  -(2784+ 3904 n_\text{s} + 1960 n_\text{s}^2 + 416 n_\text{s}^3 + 32 n_\text{s}^4 ) \widetilde{k}^7 + (370+503 n_\text{s} + 247 n_\text{s}^2 +52 n_\text{s}^3 + 4n_\text{s}^4) \widetilde{k}^8 \nonumber \\
&  - 256 ( 1 - \widetilde{k})^6 (6 + 6 (2 +n_\text{s}) \widetilde{k} + (2 + n_\text{s})(5+2 n_\text{s}) \widetilde{k}^2) \Big( 1 - \frac{ \widetilde{k}}{2 - \widetilde{k}}\Big)^{2 n_\text{s}} \Big)~,  
\end{align}

\begin{equation}
\label{eq:omega_tilt_ls2}
\widetilde\Omega_{\text{GW,RD}}^{\text{(LS,2)}}=~ 2(2 -\widetilde{k})^4 \Gamma (4+2 n_\text{s}) \left( \frac{\widetilde{k}^4}{\Gamma(5+2n_\text{s})} -\frac{4 \widetilde{k}^2 (2 -\widetilde{k})^2}{\Gamma(7+2n_\text{s})} + \frac{24(2 -\widetilde{k})^4}{\Gamma(9+2 n_\text{s})} \right)~,
\end{equation}
$\widetilde{k} = x_{\rm R}/x_{{\rm max}, {\rm R}} = k / k_{\rm max}$ and $\Gamma (x)$ is the Gamma function.
The resonance GW contribution is 
\begin{align}
\label{eq:omega_tilt_res}
\Omega_\text{GW,RD}^\text{(res)} =&~ \frac{2.30285 \times \sqrt{3} \, 3^{n_\text{s}}}{2^{13+2 n_\text{s}} \times 625 } x_\text{R}^7 \left( \frac{x_\text{R}}{x_\text{*,R}} \right)^{2(n_\text{s}-1)} s_0(x_\text{R}) \nonumber \\
& \times \left( 4 {}_2 F_1 \left( \frac{1}{2}, 1- n_\text{s} ; \frac{3}{2} ; \frac{s_0^2 (x_\text{R})}{3} \right)   -3 {}_2 F_1 \left( \frac{1}{2}, - n_\text{s} ; \frac{3}{2} ; \frac{s_0^2 (x_\text{R})}{3} \right) - s_0^2 (x_\text{R}) {}_2 F_1 \left( \frac{3}{2}, - n_\text{s} ; \frac{5}{2} ; \frac{s_0^2 (x_\text{R})}{3} \right)   \right),
\end{align}
where ${}_2 F_1 \left(a,b; c; z\right) $ is the hypergeometric
function, and $s_0 (x_\text{R})$ is given by 
\begin{align}
s_0 (x_\text{R}) =&~ \begin{cases}
1  &   x_\text{R} \leq \frac{2 x_\text{max,R}}{1+\sqrt{3}} \\
2 \frac{x_\text{max,R}}{x_\text{R}} -\sqrt{3} & \frac{2 x_\text{max,R}}{1+\sqrt{3}} \leq x_\text{R} \leq \frac{2 x_\text{max,R}}{\sqrt{3}} \\
0 &  \frac{2 x_\text{max,R}}{\sqrt{3}} \leq x_\text{R} 
\end{cases}. \label{eq:s_0}
\end{align}

\end{document}